\documentclass[aps,prfluids,reprint,superscriptaddress]{revtex4-2}

\usepackage{graphicx}
\usepackage{xcolor}

\begin{document}


\title{Stability of particle laden interfaces of drops flowing through a pore }




\author{Franz De Soete}
\affiliation{Soft Matter Sciences and Engineering (SIMM), ESPCI Paris, PSL University, Sorbonne Universit\'e, CNRS, F-75005 Paris, France}

\altaffiliation{Laboratoire Physico-Chimie des Interfaces Complexes, ESPCI Paris, 10 rue Vauquelin, F-75231 Paris, France}

\author{Nicolas Passade-Boupat}

\affiliation{TOTALEnergies S.A., P\^ole d’Etudes et de Recherches de Lacq, BP 47, 64170 Lacq, France}
\altaffiliation{Laboratoire Physico-Chimie des Interfaces Complexes, Chemstartup, RD 817, 64170 Lacq, France}

\author{Laurence Talini}
\affiliation{CNRS, Surface du Verre et Interfaces, Saint-Gobain, 93300 Aubervilliers, France}
\altaffiliation{Laboratoire Physico-Chimie des Interfaces Complexes, ESPCI Paris, 10 rue Vauquelin, F-75231 Paris, France}

\author{Fran\c cois Lequeux}
\affiliation{Soft Matter Sciences and Engineering (SIMM), ESPCI Paris, PSL University, Sorbonne Universit\'e, CNRS, F-75005 Paris, France}

\altaffiliation{Laboratoire Physico-Chimie des Interfaces Complexes, ESPCI Paris, 10 rue Vauquelin,
	F-75231 Paris, France}

\author{Emilie Verneuil}
\affiliation{Soft Matter Sciences and Engineering (SIMM), ESPCI Paris, PSL University, Sorbonne Universit\'e, CNRS, F-75005 Paris, France}\email{emilie.verneuil@espci.fr}

\altaffiliation{Laboratoire Physico-Chimie des Interfaces Complexes, ESPCI Paris, 10 rue Vauquelin,
	F-75231 Paris, France}

\email{emilie.verneuil@espci.fr}

\date{\today}

\begin{abstract}
When a drop laden with solid particles and suspended in a liquid passes
through a narrow pore, its interface experiences strong shear and
elongation, and the raft of particles may accumulate toward the back of the drop. Using well controlled formulations of Pickering drops driven at set pressure, we determine the two conditions for which solid
particles are expelled from the oil-water interface after a Pickering drop
passes a converging-diverging pore: {\it (i)} particles accumulation at
the rear of the drop is such that surface pressure builds-up at the
interface.  {\it (ii)} Surface pressure relaxation by buckling is impaired by geometrical constraints. These two conditions are
rationalized using three non dimensional numbers: the capillary number,
the particle to pore size ratio, and the drop to particle size ratio,
which allow to account for the viscous shear at the interface, the stability of the lubricating film between the pore wall and the drop,  the drag on the raft of particles adsorbed at the interface, and its mechanical behaviour.

\end{abstract}


\maketitle

\section{Introduction} \label{sec:intro}

Particle-laden drops suspended in a fluid, also called Pickering drops, bear remarkably stable interfaces : at liquid-liquid interfaces, capillarity strongly stabilizes solid particles of typical size ranging between 0.1 to 10 $\mu$m \cite{Pieranski1980, Chevalier2013}. The resulting strong adsorption makes these drops more stable than bare drops with regards to coalescence or shear \cite{gai_amphiphilic_2017, Lagarde:protiere:2020}. From a practical point of view, the behavior of suspensions of particle-laden drops in strongly sheared situations is relevant to the understanding of filtration processes \cite{Smith1985, Mehrabian2015} or oil recovery \cite{Perazzo2018, kokal2005}, where mixtures of immiscible liquids and solids are pushed through a porous medium in order to separate the two liquids. To gain insight in this complex situation, we offer to describe the behavior of a single Pickering drop flowing through a single pore, and to define the conditions for which solid particles separate from the drop. In the past, particles \cite{Sauret2017} or soft objects \cite{wyss_capillary_2010, fiddes_augmenting_2009} such as capsules \cite{leclerc_transient_2012, dawson_extreme_2015, rorai_motion_2015, luo_solute_2019, haner_sorting_2021, leopercio_deformation_2021} or vesicles \cite{fai_active_2017, park_dynamics_2020}, flowing through constrictions have been studied in order to either assess their stability \cite{dawson_extreme_2015, leopercio_deformation_2021, luo_solute_2019}, or the conditions for clogging and flow in relation to their elasticity \cite{prevot_design_2003, wyss_capillary_2010, leclerc_transient_2012, rorai_motion_2015, fai_active_2017, do_nascimento_flow_2017, oconnell_cooperative_2019, haner_sorting_2021} or to their interactions with the walls \cite{fiddes_flow_2007, fiddes_augmenting_2009, Sauret2017, park_dynamics_2020}. However, much fewer studies \cite{Mulligan2011, Liu2021} addressed the case of Pickering drops flowing through a constriction: as particle-laden interfaces bare negligible resistance to stretching compared to elastic membranes, while they both easily bend, we expect specific behaviors may emerge in the present case.

In a recent paper \cite{DeSoete2021}, we described the behavior of oil-in-water drops laden with micrometer-sized silica particles flowing through a single convergent-divergent axi-symmetrical pore. The drop was initially 5 times larger than the narrowest part of the pore, so that it strongly deformed in the pore. {The particle size was 30 times smaller than the pore. We explored a range of capillary number $Ca$ between $3.10^{-3}$ to $3.10^{-2}$, where $Ca$ compares the viscous stress at the drop interface with the capillary pressure inside the drop : $Ca=\eta V/\gamma$ with $\eta$ the outer phase viscosity, $V$ the drop velocity, and $\gamma$ its interfacial tension. In such conditions, as the drop flows through the pore, its surface expansion leads to a particle-free front interface with particles accumulating towards the rear drop interface. In the explored range of relatively small capillary numbers, we nevertheless demonstrated that the particles are efficiently driven back from the rear to the front by a Marangoni-like mechanism that opposed the viscous shear. As a consequence, the drops cross the pore without destabilization of the particle laden interface, and the drop is unchanged after its passing through the pore.
For larger capillary numbers ($10^{-2}$ to $10^{-1}$), other studies \cite{Mulligan2011} evidenced the formation of tails at the rear of the drops, with buckling and break up for highly confined drops.\\

In the present paper, we offer to determine the conditions for which solid particles are expelled from the oil-water interface after a Pickering drop passes a single pore. To do so, we explore the situations where the accumulation of solid particles at the rear of the drop does not relax. We anticipate the flow of the particle raft from the rear to the front may be slowed down if the particle size increases as compared to the thickness of the lubrication film separating the drop from the pore wall. Following Bretherton's law on bare drops or bubbles flowing through a cylinder in a liquid \cite{Bretherton1961}, the thickness $h$ of this lubrication film is expected to increase with the capillary number. Hence, we chose to vary both the adsorbed particles mean diameter and the drop capillary number in the low $Ca$ range ($10^{-4}$ to $10^{-2}$), and we investigate the consequences on both the drop movement and the flow of the particles raft adsorbed at its interface.\\

\section{Experimental system}\label{sec:exp}

\begin{figure}
	\includegraphics[width=7.5cm]{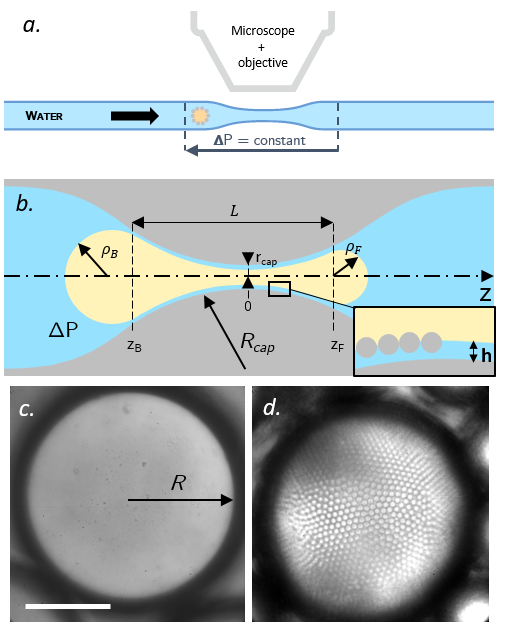}
	\caption{(a) Experimental setup : a glass cylindrical capillary tube with a narrow pore at the center is filled with water. An oil drop with radius $R$ laden with particles of size $r_s$ is pushed at constant pressure $\Delta P$ into the pore. Images are captured by a microscope and a camera. (b) Geometrical parameters of the pore and deformed drop in the pore : front and back radii of curvature $\rho_F$ and $\rho_B$, position of the drop front along the pore $z_F$. The lubrication film between the drop and the pore wall has a thickness denoted $h$. (c,d) Pickering drops laden with $r_s=375$~nm (c) and $5~\mu$m (d) silica particles. Mean drop radius is $R=125\pm25~\mu$m. Scale bar : 100~$\mu$m.} \label{fig:schema:montage}
\end{figure}

The model system we used was developed in a previous study \cite{DeSoete2021}. As model pore, we used a cylindrical tube made out of borosilicate glass with a central converging-diverging part, as depicted in Figure~\ref{fig:schema:montage}(a,b). The tube radius is $600~\mu$m away from the contraction and $r_{cap}=25~\mu$m at the constriction center. The length $L$ of the contraction is $L=3$~mm and its curvature radius is $R_{cap}=2$~mm.  The flow is driven at controlled pressure difference across the constriction. The pressure difference is denoted $\Delta P$ and varies between 1.2 and 5~kPa. The lower limit for $\Delta P$ is actually the pressure for which clogging is observed when drops do not cross the pore \cite{DeSoete2021}. The dependence between pressure difference $\Delta P$ and water flowrate $Q_w$ was carefully calibrated to measure the pore hydraulic resistance $\Psi$ defined as $\Delta P=\Psi Q_w$ for water : $\Psi=(1.4 \pm 0.1) \times 10^{12}$~Pa.s.m$^{-3}$. When compared to a computed value based on the shape of the capillary tube, this shows that only the constricted part contributes to the hydraulic resistance. The particle-laden drops are oil drops laden with silica particles and suspended in water with a background concentration of salt (NaCl at $10^{-4}$~mol.L$^{-1}$) and a small amount of cationic surfactant (CTAB at $10^{-9}$ mol.L$^{-1}$) to finely adjust the particle hydrophobicity. As oil, we used dodecane (viscosity $\eta_o=1.35~$mPa.s). The viscosity of the aqueous solution was that of pure water $\eta=0.89$~mPa.s, or, when noted, increased by addition of glycerol at 0.4~w:w ($\eta=3.12$~mPa.s) or 0.5~w:w ($\eta=5.63$~mPa.s). The interfacial tension between oil and aqueous phase was measured at $\gamma_{o/w}=38.6$~mN.m$^{-1}$ for pure water, and $35$~mN.m$^{-1}$ (resp. $33.5$~mN.m$^{-1}$) for 0.4 (resp. 0.5)~w:w water:glycerol solutions. As silica particles, we used spherical beads provided by Fiber Optic Center with a mean radius $r_s$ equal to 0.125, 0.375, 0.75, 1.5, 2.5, and 5 $\mu$m with a 7\% standard deviation. The particles are first dispersed in 10 ml NaCl solution using an ultrasonic probe (20 000 Hz, 40\% of maximum intensity). Oil is further added and the emulsion is obtained by mixing at 18 000 rpm
	for 30 s. The volume of oil is adjusted so that all particles and oil are emulsified \cite{DeSoete2021}. We obtained drops with radius $R=125~\pm 25~\mu$m and a surface coverage in particles measured at $C=0.86 \pm 0.04$. Microscopy images showing drops laden with silica particles is shown in Fig.~\ref{fig:schema:montage}(c,d). When adsorbed at the oil/water interface, silica particles with radius larger than or equal to 0.75~$\mu$m scatter light and their movement at the drop interface can be imaged, as will be detailed below. The emulsion is diluted with water so that single drops are pushed one by one in the pore. Finally, observation of pressure driven particle-laden drops through the axisymmetrical pore is made with an inverted microscope at magnification 5x and transmitted illumination. Images are acquired with a high speed camera at 10 000 to 18 000~fps. Examples of videos are available as Supplemental Material \cite{SMvideo:freeflow, SMvideo:wrinkles, SMvideo:expulsion}.\\

\begin{figure}
	\includegraphics[width=7.5cm]{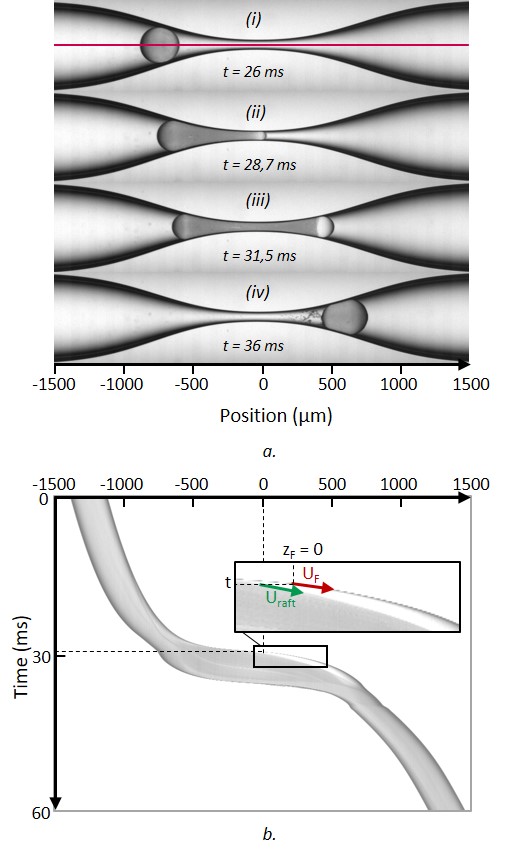}
	\caption{(a) Time series of a drop of radius $R=129~\mu$m laden with $r_s=750~$nm particles driven through a pore at $\Delta P=2700$~Pa. (b) Space-time intensity diagram along the red line in image (a,i). The darkest grey levels correspond to the particle raft adsorbed at the drop interface. The brightest area is the free-of-particles front interface. The slope of the two frontiers allows to measure the drop front velocity $U_F$ and the particle raft velocity $U_{raft}$ as shown in inset.} \label{fig:mesure:Uint}
\end{figure}

A typical time series of images is shown in Fig.~\ref{fig:mesure:Uint}a). Initial time is taken when the drop shape starts to depart from a sphere. Images ({\it ii}) and ({\it iii}) clearly evidence the surface expansion of the drop, as well as the heterogeneous surface coverage in particle, with a particle free front and particle accumulation at the rear. The particle raft is clearly delimited from the front bare part and this frontier can be tracked by image analysis. Fig.~\ref{fig:mesure:Uint}b shows a space time intensity diagram along the red line of Fig~\ref{fig:mesure:Uint}a-{\it i}). The bare front of the drop shows brighter, while the particle-laden interface corresponds to lower grey levels. This allows to measure both the front drop velocity $U_F$ and the particle raft velocity $U_{raft}$ as the slopes of the two frontiers. These velocities are shown by arrows in the inset of Fig.~\ref{fig:mesure:Uint}b in the particular case where the drop front denoted $z_F$ is at the pore center : $z_F=0$. Besides, the time series of images are used to measure the flow rate of the drop $Q$ over time, using the axial symmetry of the pore. Because the drops are driven at constant pressure difference $\Delta P$ across the contraction, we use next the classical relationship between the flow-rate $Q$, applied pressure, and capillary pressures at both the front and rear interfaces of the drop to measure the latter. To do so, three assumptions are made: the water films around the drop do not contribute to the flowrate; the pore curvature $1/R_{cap}$ can be neglected so that the flow is a Poiseuille flow; the hydraulic resistance $\Psi$ is computed from the knowledge of the pore shape $r(z)$ and the location of the oil drop \cite{DeSoete2021}. This writes:
%
%

\begin{equation}
\Delta P+\frac{2\gamma_B}{\rho_B}-\frac{2 \gamma_F}{\rho_F}=\Psi Q
\label{eq:flowrate:pressure:theo}
\end{equation}

where $\rho_F$ (resp. $\rho_B$) and $\gamma_F$ (resp. $\gamma_B$) are the front (resp. rear) radius of curvature and interfacial tension, as depicted in Fig.~\ref{fig:schema:montage}b). The radii of curvature are also measured over time by image analysis. besides, from the typical images in Fig.~\ref{fig:mesure:Uint}a) we find that, as the drop flows through the pore, the front interface is free of silica particles so that $\gamma_F=\gamma_{o/w}$, the oil/water interface tension, at all times \cite{DeSoete2021}. At the back of the drop where surface coverage in adsorbed silica particles varies, the interfacial tension $\gamma_B$ will account for the subsequent change in surface energy. By analogy with the inter-particulate pressure in 3D granular systems, when the particle raft is compressed, we offer to account for the inter-particulate forces building up between the silica beads adsorbed at the interface \cite{Liu2021} by defining a 2D surface pressure denoted $\pi$ and defined as the difference between the bare oil/water interfacial tension and the current value of the interfacial tension $\gamma$ : $\pi=\gamma_{o/w}-\gamma$. At the interface of the non deformed drop or at the bare front interface, $\pi$ is zero. If a raft of particles is compressed, at some critical surface coverage, surface pressure builds up and $\pi$ becomes positive. With this notation, the surface pressure at the front is $\pi_F=0$, and Eq.~\ref{eq:flowrate:pressure:theo} can be rewritten so as to provide a measure of the surface pressure at the back $\pi_B$:
\begin{equation}
\pi_B=\gamma_{o/w} (1-\frac{\rho_B}{\rho_F})+\frac{\rho_B}{2}\left(\Delta P -\Psi Q\right)
\label{eq:pi:measure}
\end{equation}
Here, assumption is made that as long as the back of the drop is hemispherical, its surface pressure is isotropic.\\
In the following, we offer to discuss the various regimes adopted by a drop flowing through a converging-diverging pore depending on its velocity and the silica particle size, in terms of capillary number defined as :
\begin{equation}
Ca=\frac{\eta U_F}{\gamma_{o/w}}
\label{eq:def:Ca}
\end{equation}
and surface pressure $\pi_B$ at the rear interface of the drop.

\section{Results}

\begin{figure*}
	\includegraphics[width=15cm]{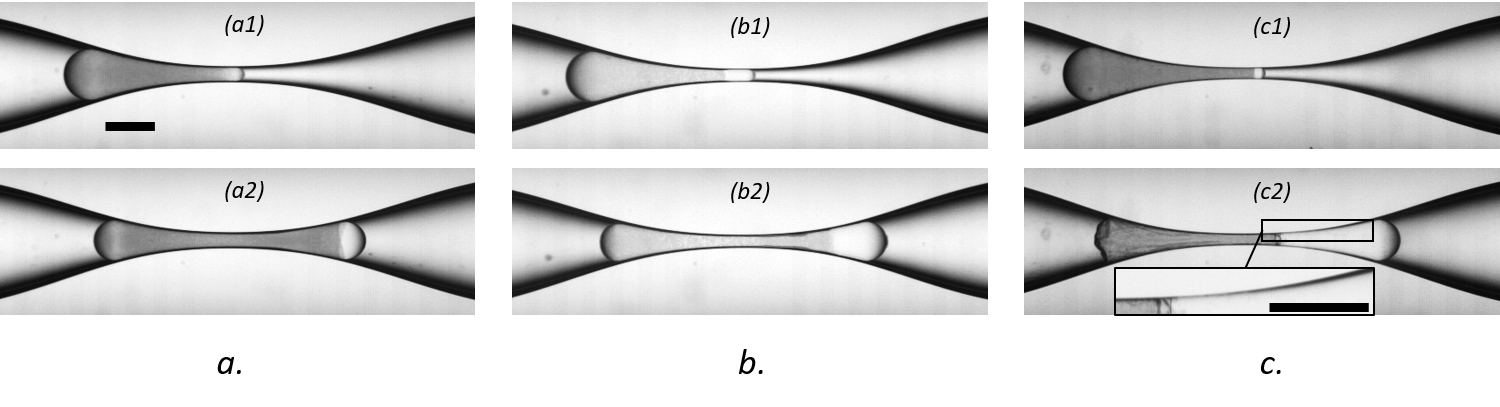}
	\caption{Flow regimes of the particle raft adsorbed at the drop interface depending on the capillary number Ca and particle size $r_s$. (a) $r_s$=0.75~$\mu$m, Ca=10$^{-2}$. (b) $r_s$=2.5~$\mu$m, Ca=10$^{-2}$. (c) $r_s$=0.75~$\mu$m, Ca=10$^{-4}$. The raft always accumulates at the back. In (a), it relaxes towards the front and only the front apex of the drop remains naked. See also video \cite{SMvideo:freeflow}. (b) The naked front part extends over a longer part of the drop. (c) An even longer naked front zone is observed, where oil wets the pore wall (inset - see also \cite{SMvideo:wrinkles}). Scale bar : 100$\mu$m. } \label{fig:regimes: accumulation}
\end{figure*}

As introduced in Section~\ref{sec:intro}, when the radius of the adsorbed silica particles compares with the thickness of the lubrication film squeezed between the raft and the pore wall, the movement of the raft is expected to be all the more difficult than the particle size is closer to the film thickness. The latter can be derived assuming that the lubrication film has a thickness denoted $h$ set by the Bretherton's law \cite{Bretherton1961}, which applies to particle-free drops driven in a cylinder within a fluid. The cylinder radius is taken at the center of the pore. This yields :
\begin{equation}
h=1.34 r_{cap} Ca^{2/3}
\label{eq:bretherton}
\end{equation}
With $r_{cap}=25$~$\mu$m and $Ca$ ranging between $10^{-4}$ and $10^{-2}$, the lubricating film thickness is expected to vary in the 0.1 to 3 $\mu$m-range. Note that more refine laws have been derived in the literature that account for the viscosity ratio of the outer and inner phases \cite{Balestra:2018}, inertial effects \cite{Aussilous2000}, or adsorbed particles \cite{stone_2017}. In the present range of capillary number, with a viscosity ratio of 3, the lubrication film thickness would only differ by a factor of order one and less than 2 from Bretherton's model. Since both the capillary radius and the surface coverage in particles change within the constriction, we choose to ignore these corrections. We will see later that Bretherton's hypothesis is sufficient to describe the destabilization mechanisms of the drops.\\
Hence, we first set the lubrication film thickness by selecting experiments with roughly the same capillary number, and we increase the silica particle size. Images are shown in Figure~\ref{fig:regimes: accumulation}a) and b) where the capillary number Ca is 10$^{-2}$, which would correspond to $h\sim 1.5~\mu$m through Eq.~\ref{eq:bretherton}, and particle radius increases from 0.75~$\mu$m to 2.5~$\mu$m between a) and b). Images in each row correspond to the same drop position, and the two rows to two subsequent times. Comparison of images (a1) with (b1) or (a2) with (b2) shows that the particle raft is all the more accumulated towards the rear that the silica particles are large : the surface coverage at the rear increases with $r_s$ at a given Ca number, and thereby, surface pressure probably builds up. Next, we offer to decrease the capillary number with respect to the case shown in Fig.~\ref{fig:regimes: accumulation}a), for the same silica particle radius of $r_s=0.75~\mu$m. In Figure~\ref{fig:regimes: accumulation}c), Ca=$10^{-4}$. The bare front of the drop now extends over an even larger area, the particle raft being so confined towards the back that the rear of the drop crumples and wrinkles appear along the drop in Fig.~\ref{fig:regimes: accumulation}c2). Besides, a careful analysis of the grey levels on the inset of Image~\ref{fig:regimes: accumulation}c2) evidences that the bare front part of the drop is no longer separated from the pore wall by a lubricating water film : oil wets the wall. This is clearer on the movie available as Supplemental Material \cite{SMvideo:wrinkles}. Quantitatively, we find that the velocity of the particle raft $U_{raft}$ drops to zero as the front passes the contraction center. In Figure~\ref{fig:Uint:Uf}a, we systematically measure the variations of the particle front velocity $U_{raft}$ as a function of the drop capillary number $Ca$ (Eq.~\ref{eq:def:Ca}) when the drop front is at the pore center. This position corresponds to the first row of Figure~\ref{fig:regimes: accumulation}. Here, the particle radius is $r_s=0.75~\mu$m. We find that below Ca$^*=6.10^{-4}$ - which corresponds in this case to $U_F^*=0.025$~m/s, the particle raft velocity is zero and oil wets the pore wall. Above $Ca^*$, the particle velocity is always one order of magnitude lower than the drop velocity. We also systematically ompute the surface pressure at the back of the drop $\pi_B$ through Eq.~\ref{eq:pi:measure} as a function of $Ca$ in Figure~\ref{fig:Uint:Uf}b). As anticipated, we find that as the particle raft velocity decreases, particles accumulates at the back and simultaneously, surface pressure builds up. \\

\begin{figure}
	\includegraphics[width=8cm]{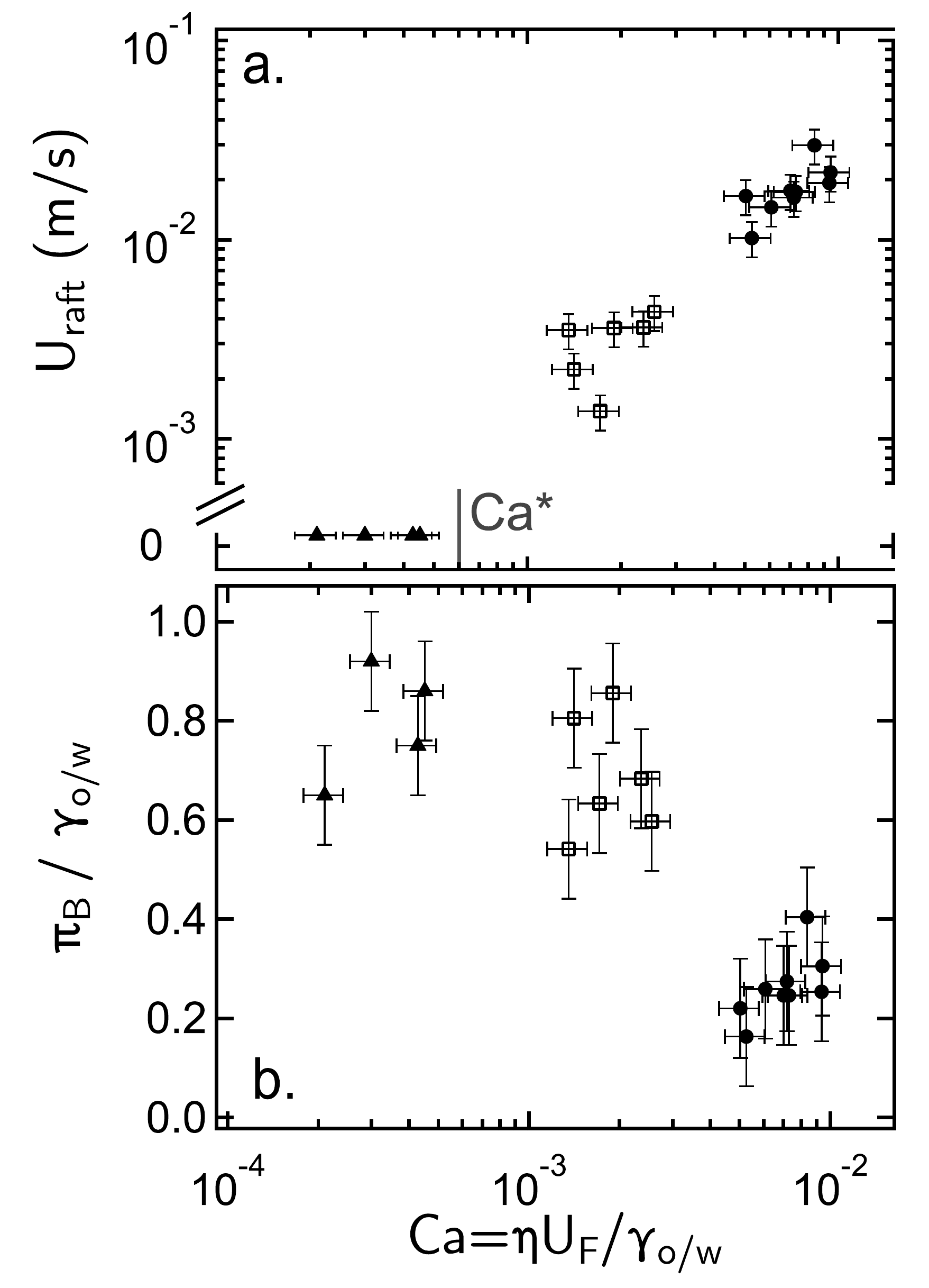} 
	\caption{(a) Interfacial velocity $U_{raft}$ and (b) surface pressure at the drop back $\pi_B$ as a function of the capillary number Ca defined with the front velocity $U_F$ for Pickering drops laden with 750 nm radius silica particles. Full triangles : wetting of oil on the pore wall is observed.} \label{fig:Uint:Uf}
\end{figure}

We chose next to explore in details the low capillary number case, and we vary the particle radius over the whole available range. In Figure~\ref{fig:images:wetting:lowCa:rs}, the capillary number is set around Ca$\sim10^{-4}$ and each column corresponds to a different particle radius : $r_s$ is 0.125$~\mu$m in Fig.\ref{fig:images:wetting:lowCa:rs}a), and 5~$\mu$m in (b). Each row corresponds to the same drop position, and the four rows are four subsequent times or positions. In all cases, the particle raft is stopped, oil wets the wall at the front of the drop - as indicated by the contrast change in inset in Fig.~\ref{fig:images:wetting:lowCa:rs}a2) and in the video \cite{SMvideo:wrinkles} -, and particles are forced to accumulate at the drop rear. This accumulation at the rear changes the behavior of the rear interface of the drop depending on $r_s$. In Figs.\ref{fig:images:wetting:lowCa:rs}a2, longitudinal wrinkles appear at the rear of the particle raft (see also video \cite{SMvideo:wrinkles}), whereas no wrinkles are detected in Fig.\ref{fig:images:wetting:lowCa:rs}b2. At later times, particles do not separate from the liquid drop in a4, and the Pickering drop, remarkably, crosses the pore with no damage, at variance with case (b) where the particles fully separate from the drop interface : in (b4), after the drop has crossed the pore, its interface is naked and silica particles are now dispersed in the aqueous phase (see also video \cite{SMvideo:expulsion}).\\

\begin{figure*}
	\includegraphics[width=15cm]{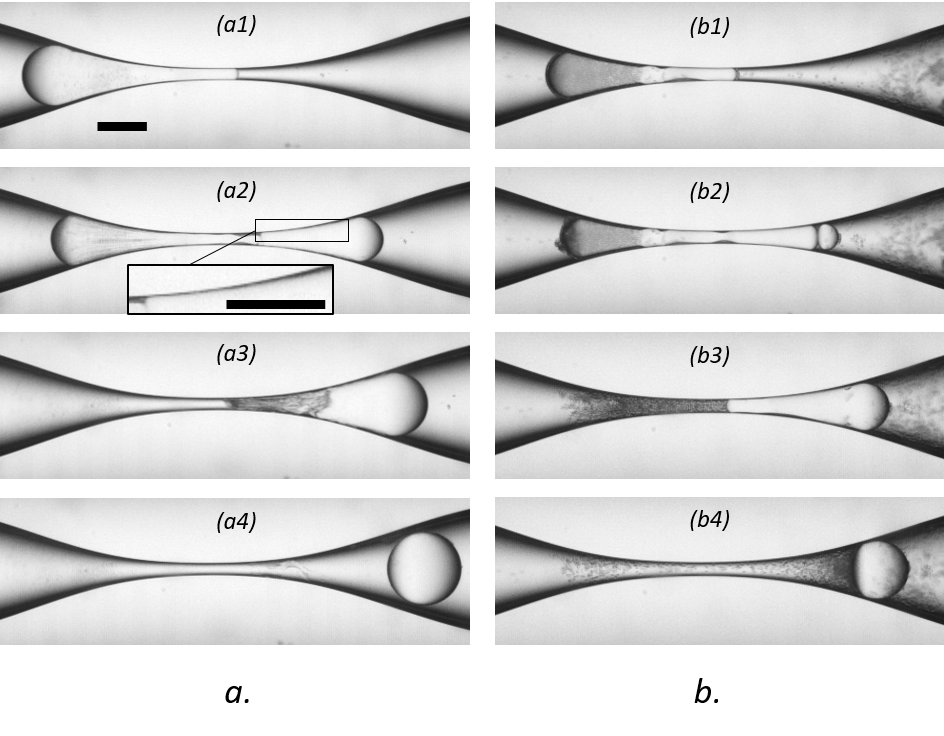}
	\caption{ Effect of the particle radius at low capillary number on the interface compression and relaxation : Ca$\sim$10$^{-4}$. (a) $r_s$=0.125~$\mu$m, (b) $r_s$=5~$\mu$m. Each column presents 4 subsequent images as a drop passes through the pore. Each line corresponds to the same position of the drop. The oil in the drop wets the pore wall at the front (inset) where the oil/water interface is naked. At the back, where the raft is adsorbed at the interface, the lubrication water film persists.  (a)  Wrinkling in a2. No particle expulsion.  (b) No wrinkling. Full particle expulsion (b4). Scale bars : 100 $\mu$m.
	} \label{fig:images:wetting:lowCa:rs}
\end{figure*}

Similar behaviors were obtained by systematically varying the applied pressure, and thus the drop velocity, and the particle radius $r_s$. The results are summarized in the $r_s$ versus Ca diagram of Figure~\ref{fig:rs:Ca}.

\begin{figure}
	\includegraphics[width=8cm]{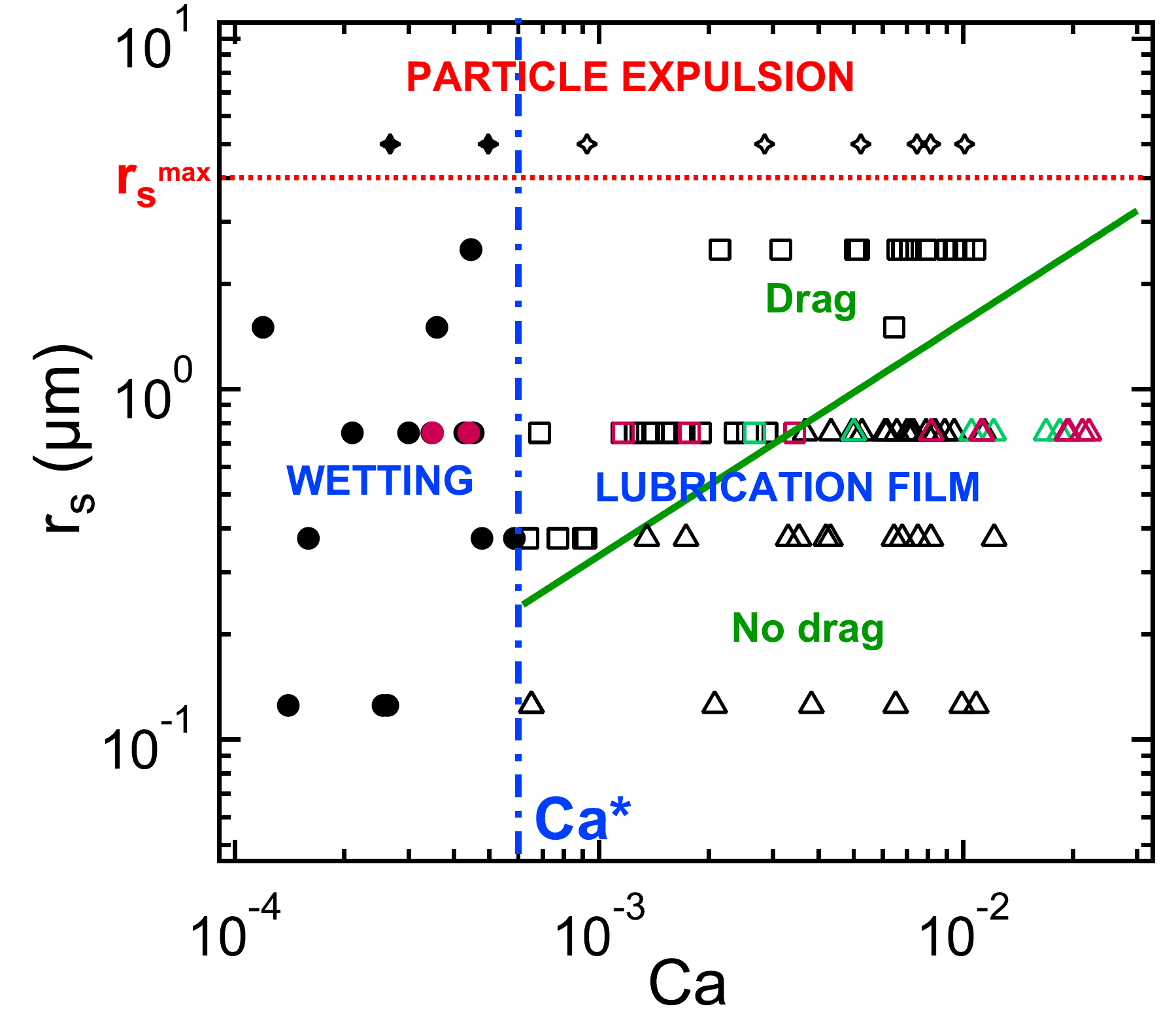}
	\caption{Flow regimes for a Pickering drop covered with particles of radius $r_s$ flowing through a pore at capillary number Ca. Aqueous phase: black symbols: water; red: water/glycerol 0.4 w:w; green: 0.5 w:w. Hollow symbols: a lubricating water film separates the oil drop from the pore wall. Full symbols: the lubricating water film dewets and oil wets the wall. Blue dash-dotted line: frontier Ca=Ca$^*$. Squares: drag $\Delta \sigma >0$ (see Eq.~\ref{eq:delta:sigma}). Triangles: no drag, $\Delta \sigma =0$, no wrinkles, no expulsion.  Green line: $r_s=h$ (Eq.\ref{eq:limit:bretherton}). Circles: wrinkles observed at the drop back. Diamonds: no wrinkles, full particles expulsion. Red dotted line: $r_s=r_s^{max}=R/A=4~\mu$m. }
	\label{fig:rs:Ca}
\end{figure}

\section{Discussion}

\subsection{Movement of the particle raft at the drop interface}

In Figure~\ref{fig:rs:Ca}, the wetting cases for which $U_{raft}=0$ are shown as full symbols while the non wetting cases are hollow symbols. We clearly find that the frontier between wetting and non wetting corresponds to a defined value of the capillary number, denoted $Ca^*$. This was further checked by changing the aqueous phase viscosity with glycerol addition (green and red markers) which allows to vary $Ca$ at constant drop velocity. In Figure~\ref{fig:rs:Ca}, a dash-dotted line marks the frontiers between the wetting and non wetting cases, and we find Ca$^*=6.10^{-4}$. In the following, we discuss this value in terms of the condition for which the water lubricating film dewets, and the velocity at which, once nucleated, the oil wets the pore wall. Our observations show that oil wets the wall where the drop interface is naked. Once nucleated, the wetting of oil propagates towards the front part of the drop, the propagation towards the rear being impaired by the silica particles. Therefore, we offer to compare the critical value Ca$^*$ we measure to the spontaneous wetting velocity of oil on silica and within a submicrometer thick water film for the same system (dodecane and CTAB with same concentration), taken from a previous work \cite{Rondepierre2021}: we had found that wetting was easily nucleated and that oil spontaneously wetted silica in the water film at a capillary number Ca$_s$=3.10$^{-4}$ which was found independent on the film thickness, but system-dependent. Here, we find a good agreement with the present capillary number value Ca$^*$. We conclude that as soon as the drop capillary number falls below the spontaneous oil wetting value $Ca_s$, nucleation of oil wetting occurs and propagates faster than the drop moves. As a consequence, the particle raft stops and $U_{raft}$=0. \\

\begin{figure}
	\includegraphics[width=8cm]{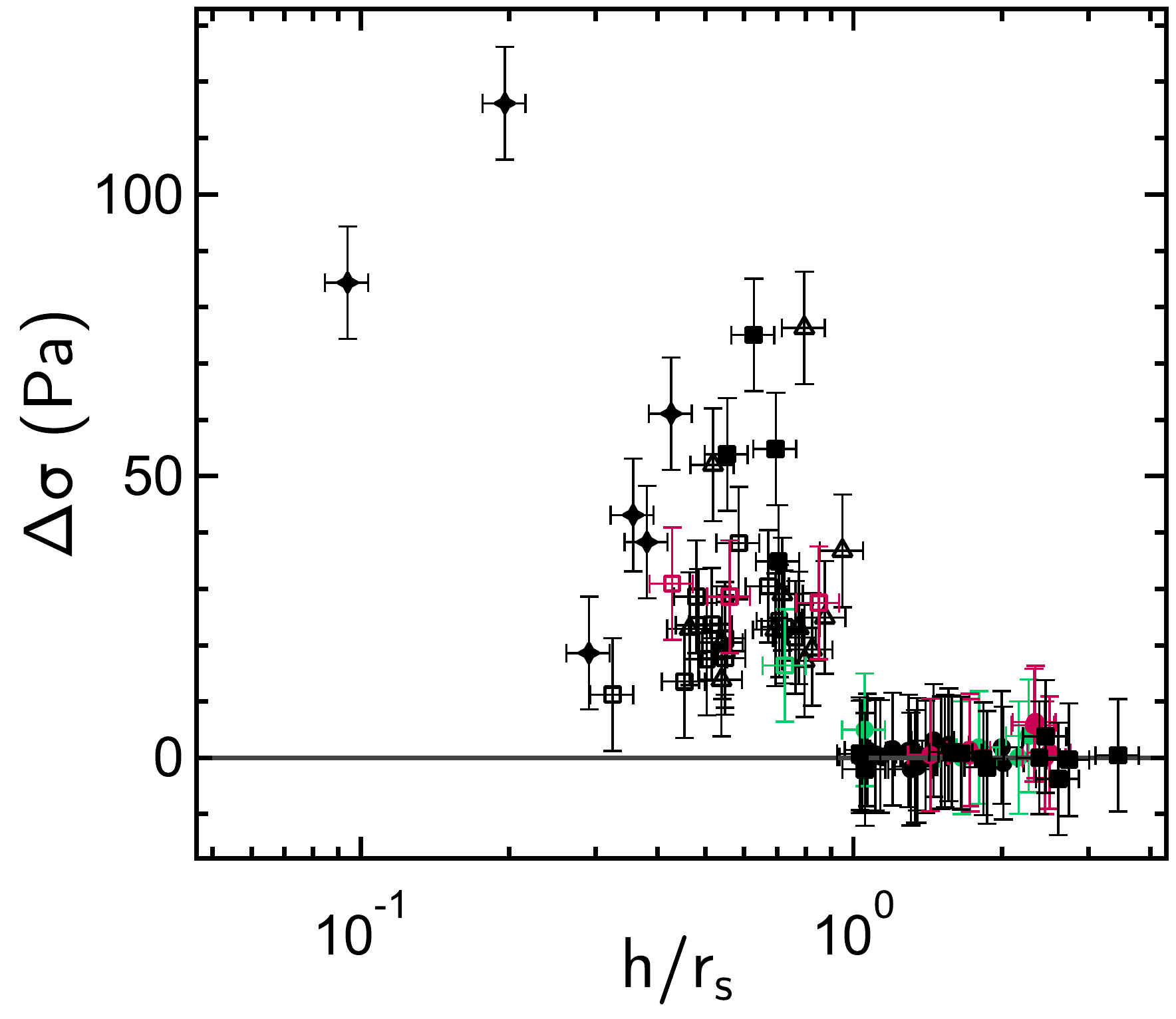}
	\caption{Additional drag acting on the particle raft as defined by Equation~\ref{eq:delta:sigma} as a function of the lubrication film thickness $h$ over particle radius $r_s$ ratio for all the experiments of Fig.~\ref{fig:rs:Ca} (same symbols). Thickness $h$ computed from Eq.~\ref{eq:bretherton}. $\Delta \sigma$ decreases to zero for $h/r_s\sim 1$. Aqueous phase: black: water; red: water/glycerol 0.4 w:w; green: 0.5 w:w.} \label{fig:delta:sigma}
\end{figure}

Beyond the wetting case for which silica particles obviously accumulate towards the back of the drop because the raft velocity drops to zero, as in Fig.~\ref{fig:regimes: accumulation}c, experiments at Ca$>$Ca$^*$ show that the raft velocity decreases when the particle radius increases (see Fig.~\ref{fig:regimes: accumulation}a-b). In the following, we derive the mechanical balance accounting for the particle raft movement towards the front of the drop. We first examine the cases where 
capillary numbers are larger than Ca$^*$ and particle are small (typically $r_s=0.75~\mu$m), as in the video presented in \cite{SMvideo:freeflow}. The raft movement from the rear to the front of the drop was described in a previous work \cite{DeSoete2021} to be simply driven by the surface pressure gradient $\Delta \pi/L$ between the back and front of the drop of length $L$, and opposed by the viscous drag within the lubrication water film. For a lubricating film of thickness denoted $h$, the viscous drag writes $\eta \frac{U_{raft}}{h}$. \\

In the present paper, we assume that the lubrication film has a thickness set by Bretherton's law Eq.~\ref{eq:bretherton}, so that increasing the particle size at constant Ca or decreasing Ca could result in an additional drag on the raft arising when the particle size becomes as large as the lubrication film thickness. To test this hypothesis, we define the additional drag $\Delta \sigma$ acting on the particle raft as the difference between the driving term $\Delta \pi/L$ and the viscous drag. With our notations where $\pi_F$ and $\pi_B$ are the surface pressures at the front and back of the drop respectively, we have $\Delta \pi=\pi_B-\pi_F$ and $\pi_F=0$, so that the additional drag writes:
\begin{equation}
\Delta \sigma = \frac{\pi_B}{L}-\eta \frac{U_{raft}}{h}
\label{eq:delta:sigma}
\end{equation} 
This additional drag is a force per unit surface acting on the particle raft which was systematically measured for drops with varied silica particle sizes and varied drop velocities, and the result is plotted in Figure~\ref{fig:delta:sigma} as a function of the ratio between the silica particle size $r_s$ and the Bretherton's thickness $h$. We find that $h/r_s$=1 clearly marks the limit between the case where the compressed raft freely relaxes under the combined effects of the surface pressure gradient and the viscous drag ($\Delta\sigma=0$), and the cases where $\Delta\sigma$ is non zero. This observation confirms the onset of an additional drag when the lubrication film thickness, as computed from Bretherton's law, compares with the particle size. This additional drag can be thought of as a friction term of the silica particles sliding against the pore wall as soon as $h<r_s$. This result allows to refine the diagram in Figure~\ref{fig:rs:Ca} : experiments for which $\Delta \sigma=0$ are plotted as circles, and $\Delta \sigma \ne 0$ as squares. The line corresponds to the frontier set by the condition $h=r_s$ which writes, through Equation~\ref{eq:bretherton} :
\begin{equation}
r_s=1.34 r_{cap} Ca^{2/3}
\label{eq:limit:bretherton}
\end{equation}
\\

As a first result, we rationalize the existence of a region of the $r_s$ versus $Ca$ diagram where the particle accumulation at the back of the drop relaxes through the sole dissipative effect of the viscous drag : this region is delimited at low Ca by $Ca^*$ given by the spontaneous wetting velocity of oil on silica in water, and at large radius of particles by the water film thickness as calculated by Bretherton's law. It corresponds to the hollow circle markers in Fig.~\ref{fig:rs:Ca}.\\
Outside this region, particles adsorbed at the drop interface accumulate at the back of the drop and surface pressure builds up. In the following, we examine the consequence of an excess of surface pressure in the regions where either friction or wetting impair the surface pressure relaxation. These regions are marked in Fig.~\ref{fig:rs:Ca} as triangles, squares and diamonds. In particular, we will link the surface pressure excess to both the wrinkles and the particle expulsion we observed in Fig.~\ref{fig:images:wetting:lowCa:rs}).\\

\subsection{Mechanical destabilization of the particle raft at large surface pressure}\label{sec:mech:2D}

\begin{figure*}
	\includegraphics{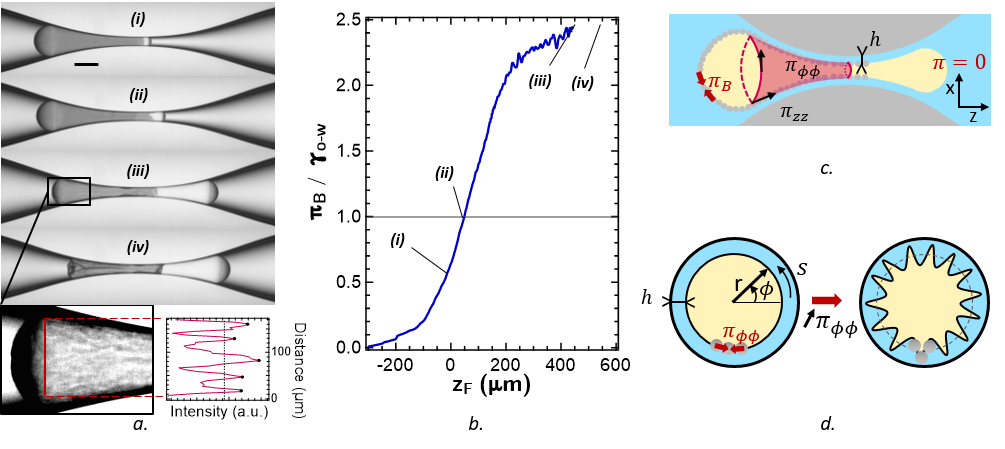}
	\caption{(a) Time series of images of a drop laden with $r_s=750 nm$ particle entering the pore at Ca$=2.10^{-4}$. $z_F=$ {\it (i)} 0; {\it (ii)} 100~$\mu$m; {\it (iii)} 465~$\mu$m; {\it (iv)} 550~$\mu$m. Inset in {\it (iii)} shows the onset of longitudinal wrinkles of wavelength $\lambda=35~\mu$m. (b) Normalized surface pressure $\pi_B$ at the back of the drop versus drop front position $z_F$. The onset of wrinkles corresponds to $\pi_B=2.4 \gamma_{o/w}$. (c) Schematic of the surface pressures at the drop interface at the onset on wrinkling : at the frontier between the back spherical cap and the wrinkled interface (red dotted line), an orthoradial component of surface pressure $\phi_{\phi\phi}$ builds up due to the section reduction. (d) Schematical views of the drop cross section separated from the pore by a lubricating film of thickness $h$ : increasing the orthoradial surface pressure $\pi_{\phi\phi}$ causes the buckling of the interface. Cylindrical coordinates ($r$, $\phi$). Curvilinear coordinate $s=r\phi/2\pi$.
	}
	\label{fig:wrinkles:images:pression}
\end{figure*}

We first examine the wrinkling of the particle-laden interface at the back of the drop. Such phenomenon was reported in the past for capsules in a pore \cite{dawson_extreme_2015} or shear particle-laden drops \cite{Liu2021}, although its onset condition was not characterized in the latter case. Figure~\ref{fig:wrinkles:images:pression}a displays a series of images of a drop advancing through a pore in a wetting case where the particle raft stops : $U_{raft}=0$, and the particle-laden interface wrinkles as shown in the inset of Fig.~\ref{fig:wrinkles:images:pression}a-iii) which corresponds to the time when wrinkles appear. The wrinkles are aligned with the longitudinal direction, and develop where the interface is squeezed by the converging shape of the pore. At this location, the decrease of the pore section is assumed to result in an anisotropy of the surface pressure with a longitudinal component denoted $\pi_{zz}$ and an orthoradial component denoted $\pi_{\phi\phi}$. This assumption is supported by previous works \cite{Jambon2017} in which rafts of particles adsorbed at liquid/liquid interfaces were shown to exhibit an elasto-plastic behavior due to both cohesion between particles and friction between contacting particles when surface concentration becomes large enough. 
As depicted in Fig.~\ref{fig:wrinkles:images:pression}c), we first assume a continuity of the longitudinal component of the surface pressure between the hemispherical cap of the drop back and the converging part where the wrinkles appear, so that $\pi_{zz}=\pi_B$ at the frontier between the wrinkled part and the cap. 
In Figure~\ref{fig:wrinkles:images:pression}b), we measure the surface pressure of the cap at the back of the drop, $\pi_B$, as a function of the position of the drop when it flows though the pore. 
The data correspond to the image series in Fig.~\ref{fig:wrinkles:images:pression}a). Note that at times later than image {\it iii}, the drop back is no longer hemispherical, and the hypothesis of an isotropic pressure at the back fails so that we no longer compute values for the surface pressure. Wrinkling onsets in Image~{\it iii} for which the longitudinal pressure is denoted $\pi_{zz}^w$. Its averaged value over all our experiments is measured at $\pi_{zz}^w=(2.4 \pm 0.1) \gamma_{o/w}$, far larger than the expected threshold for buckling found in the literature to be $\gamma_{o/w}$ \cite{Garbin2019, Pitois2019}. We also observe that the wrinkles develop along the longitudinal direction. From these two observations, we understand that {\it (i)} the build-up of the orthoradial component of the surface pressure $\pi_{\phi\phi}$ is the only component responsible for the mechanical buckling of the particle-laden interface as depicted in Figure~\ref{fig:wrinkles:images:pression}d), and {\it (ii)} the compressed raft adopts an elastoplastic behavior: large strains caused by the section reduction plasticize the raft. Assumption is further made that the wrinkling involves small enough strains to be described within the elastic framework, which is supported by noting that buckling releases the stress and does not lead any additional plastic deformation.\\

Hence, following past studies on the buckling of particle rafts at liquid-fluid interfaces \cite{Vella2004, Protiere:2017}, we explore the onset of such a buckling instability within the framework of a plate with bending modulus $B$ supported by a thin viscous film of thickness $h$ and viscosity $\eta$. This situation is depicted in Figure~\ref{fig:wrinkles:schema:lambda}a). While the compressive stress $\pi_{\phi\phi}$ drives the development of the interface deformation with amplitude $\delta h$ and wavelength $\lambda$, the viscous dissipation within the lubricating film prevents its development. \\
The modeling of the buckling of the particle laden interface supported by a thin viscous film is detailed in Appendix~\ref{app:lambda}. It allows for the prediction of the wrinkling wavelength $\lambda$ (Eq.~\ref{eq:theo:lambda}) that depends linearly on the particle radius $r_s$ through $\lambda=A.r_s$ where $A$ depends on the orthoradial pressure $\pi_{\phi\phi}$ through $A=4\pi \left[8(1-C)(1+\nu)(\pi_{\phi\phi}/\gamma_{o/w}-1)\right]^{-1/2}$.

\begin{figure}
	\includegraphics[width=8cm]{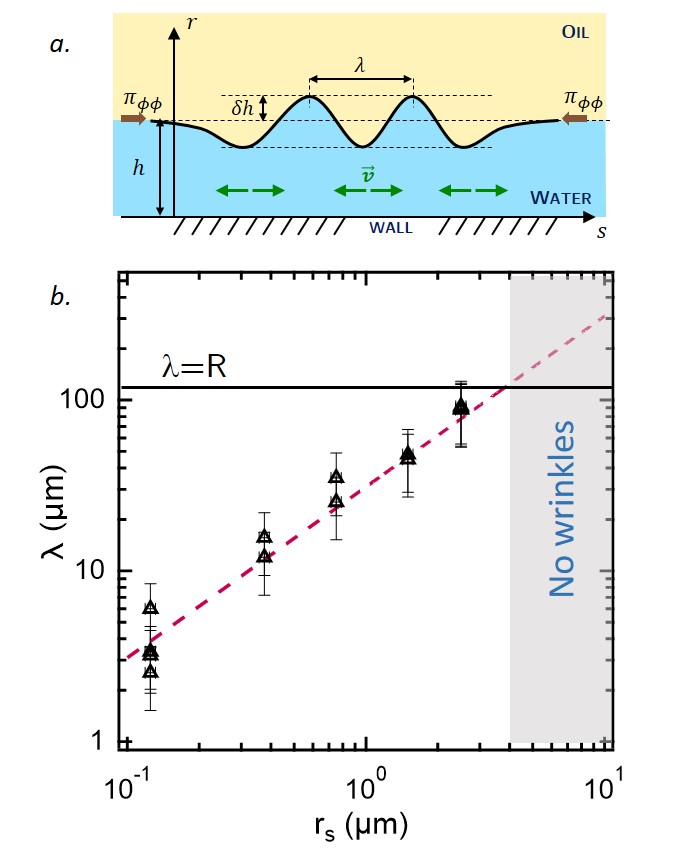}
	\caption{(a) Schematic representation of the interface buckling driven by compressive stress $\pi_{\phi\phi}$. Curvilinear abscissa $s$ defined in Fig.~\ref{fig:wrinkles:images:pression}. The deformation with wavelength $\lambda$ and amplitude $\delta h$ induces flows in the water film of mean thickness $h$ with velocity field $v$ which are limited by viscous dissipation. (b) Experimental wavelength $\lambda$ of the wrinkles at the onset of buckling as a function of particle radius $r_s$ for drops of radius $R=125~\mu$m entering the pore at varied low Ca. Black line : $\lambda=R$. Dotted line : Equation~\ref{eq:theo:lambda} with fitted prefactor $A=30$. Grey zone : no wrinkles are observed experimentally.} \label{fig:wrinkles:schema:lambda}
\end{figure}

This prediction Eq.~\ref{eq:theo:lambda} is compared with our data in Figure~\ref{fig:wrinkles:schema:lambda}b), where the wavelength $\lambda$ of the wrinkles at birth is measured by image analysis as shown in Figure~\ref{fig:wrinkles:images:pression}a). The results are plotted in Figure~\ref{fig:wrinkles:schema:lambda}b) as a function of $r_s$. We find that the wavelength increases linearly with the particle size as predicted : $\lambda=A.r_s$. From the slope of this line measured at $A\sim 30$, we measure the surface pressure $\pi_{\phi\phi}^w$ at the onset of wrinkling: $\pi_{\phi\phi}^w\sim (1.1 \pm 0.1)\gamma_{o/w} $ using $C=0.86$ as measured in Section~\ref{sec:exp} and $\nu=1/\sqrt{3}$ \cite{Vella2004}. At this stage, two comments can be made on this threshold value of surface pressure. First, it is of the order but larger than $\gamma_{o/w}$, in agreement with buckling experiments on particle-laden interfaces \cite{Garbin2019, Pitois2019}. Second, at onset of wrinkling, the compressed particle-laden interface can be regarded as a two dimensional cohesive granular material.
Following the path of previous works on both cohesive wet granular media \cite{Andreotti2013} and elasto-plasticity of compressed rafts \cite{Jambon2017}, we derive in Appendix~\ref{app:mohrcoulomb} a relationship between the two components of the surface pressure when the plasticity threshold is reached using a Mohr-Coulomb criterion that writes :
\begin{equation}
\pi_{zz}\frac{1-\sin\delta}{1+\sin\delta}-\pi_{\phi\phi}= 2c\frac{\cos\delta}{1+\sin\delta}
\label{eq:pi:phi:phi}
\end{equation}
where $\delta$ is the friction angle between silica particles and $c$ the cohesion of the raft. From the literature on silica particles sliding against glass at pH=6, $\delta=20^o$. At the onset of wrinkling, $\pi_{\phi\phi}^w=1.1\gamma_{o/w}$ and $\pi_{zz}^w=2.4\gamma_{o/w}$, so that Eq.\ref{eq:pi:phi:phi} provides an estimate of the cohesion: $c=(0.07\pm 0.02) \gamma_{o/w}$. Hence, we find a non-zero value for the cohesion, which is in agreement with our observations that particles tend to self-assemble into rafts. Our particles are small enough for gravity to be neglected, so that cohesion is likely to originate from capillary attractions due to the pinning of the oil/water contact line on the silica particles \cite{Danov:2010}. This pinning has been thought to induce non circular contact lines which generate multipolar capillary interactions between particles. The magnitude of these capillary interactions are expected to be of the order of the interfacial tension, and are found here to be around one tenth of it. In Apendix~\ref{app:mohrcoulomb}, we further derive a relationship between the cohesion $c$ and a geometrical parameter $\epsilon$ characterizing the non-circularity of the oil/water/particles contact lines that we evaluate.\\ 

From this, the following picture emerges : at the wrinkling threshold, the stress state of the 2D cohesive granular medium is set by large deformations arising from the squeezing of the drop section within the converging pore, and the plasticity threshold is reached in which frictional contacts build up between particles. Nevertheless, the strains at stake in the development of the instability are small and can therefore be described within the elastic approximation. \\

\subsection{Particle expulsion from the drop interface after crossing the pore}

From this description of the stress relaxation at the drop interface through wrinkling, we show next how to predict the conditions for drops destabilization and particle expulsion.
In Figure~\ref{fig:wrinkles:schema:lambda}b), we indicate as a grey zone the experimental conditions for which no wrinkles are observed when the drop enters the convergent part of the pore, which also corresponds to situations where full particle expulsion from the drop is observed at the pore exit. First, these two observations can be linked by noting that no wrinkling leads to no release of the compressive stress within the particle raft, until particle expulsion occurs. Second, the threshold between wrinkling and no wrinkling can be thought of as the limit where the buckling wavelength becomes larger than the drop radius. This writes : $\lambda = R$ or equivalently $A.r_s=R$ which sets a particle size limit over which expulsion from the drop interface occurs : $r_s^{max} = R/A \sim 4~\mu$m in our experimental conditions. The condition $\lambda=R$ has been reported as a line in Figure~\ref{fig:wrinkles:schema:lambda}b) and $r=r_s^{max}$ as a dotted line in the regime diagram of Figure~\ref{fig:rs:Ca}. We find that this prediction agrees with our experimental observation of full particle expulsion. Altogether, we demonstrate that the drop interface destabilization proceeds through the combination of two effects : {\it(i)} Surface pressure build-up at the back of the drop by particle accumulation in the raft; {\it (ii)} No release of surface pressure through either particle movement towards the front or buckling of the interface.\\

\section{Conclusion}

We demonstrate that particle expulsion from the rear interface of particle-laden drops flowing through a converging-diverging pore happens when two conditions are simultaneously met : {\it (i)} particles accumulation at the rear of the drop is such that surface pressure builds-up at the interface.  {\it (ii)} Surface pressure relaxation by buckling is impaired by geometrical constraints. \\
Quantitatively, we find that condition {\it (i)} amounts to a constraint on the capillary number and the particle to pore size ratio : Condition {\it (i)} is indeed met when the lubrication water film squeezed between the oil drop and the pore wall is thinner than the particle size or breaks up. In terms of capillary numbers, this writes: Ca$<(\frac{r_s}{r_{cap}})^{3/2}$ or
Ca$<$Ca$^*$ where $Ca^*$ characterizes the spontaneous wetting velocity of the inner phase (oil) on the pore wall within a water film and is system-dependent.\\
Condition {\it (ii)} constrains the drop to particle size ratio and writes : $R/r_s<A$ where $A\sim 30$ and decreases if the initial surface coverage in particles initially adsorbed at the interface decreases. \\
In line with past studies where the single pore case was extended to porous media \cite{Sauret2017, Perazzo2018, 		oconnell_cooperative_2019, Benet:2018}, we anticipate our results could efficiently be applied to the control of processes where Pickering emulsions flow through porous media, such as filtering of mixtures of immiscible liquids and solids.

\appendix
\section{Modeling the buckling of a particle laden interface supported by a thin viscous film }\label{app:lambda}

In the following, we model the growth of a sinusoidal deformation $\tilde h(s,t)=h+\delta h e^{\omega t +2i\pi qs}$ of growth rate $\omega$ and wave vector $q=2\pi/\lambda$ of the particle-laden interface.\\
First, the compressive and bending moduli are taken from the literature \cite{Vella2004} on the elasticity of interfacial particle rafts: $E\sim \frac{1-\nu}{1-C}\frac{\gamma_{o/w}}{2r_s}$ and $B=\frac{2}{3(1-\nu^2)}Er_s^3$ where $\nu=1/\sqrt 3$ \cite{Vella2004} is the Poisson ratio.\\

Second, the pressure field in the lubricating film is derived as a function of the curvilinear distance $s$ along the orthoradial direction. The pressure $P(s,t)$ obeys both a mechanical balance at the interface and the Navier-Stokes equation within the lubricating film. The fluid velocity in the orthoradial direction that develops as the interface deforms is denoted $v(s,t)$. It is averaged over the thickness of the film and varies with space and time with the same wavelength and growth rate as $\tilde h$ : $v=\delta v e^{\omega t +2i\pi qs}$. With these notations, the mechanical and hydrodynamical balance equations write:
\begin{eqnarray}
P=(\pi_{\phi\phi}-\gamma_{o/w})\frac{\partial^2 \tilde h}{\partial s^2}+B\frac{\partial^4 \tilde h}{\partial s^4}            \\
\frac{\partial P }{\partial s}+\eta_w \frac{\partial^2 v}{\partial r^2} =0
\label{eq:mech:wrinkles}
\end{eqnarray}  

The two equations~\ref{eq:mech:wrinkles} are combined into :
\begin{eqnarray}
(\pi_{\phi\phi}-\gamma)\frac{\partial ^3\tilde h}{\partial s^3}+B \frac{\partial ^5\tilde h}{\partial s^5}=\eta_e\frac{v}{h^2}\\
\end{eqnarray}

These mechanical equations are supplemented with a volume conservation equation in the fluid : 
\begin{equation}
\frac{\partial \tilde h}{\partial t}+\frac{\partial v \tilde h}{\partial s}=0
\label{eq:mass:cons:wrinkles}
\end{equation}

For the particular case of a sinusoidal deformation of the interface, the problem reduces to :
\begin{equation}
\omega=\frac{h^3}{\eta_e}q^4*(\pi_{\phi\phi}-\gamma_{o/w}-Bq^2)
\end{equation}
Following classical descriptions of instability growth, the selected wavelength is that with the maximum growth rate and is thus given by $\frac{d \omega}{dq}=0$. This condition leads to a sine deformation of the drop interface of wavelength $\lambda$ that depends on the orthoradial pressure $\pi_{\phi\phi}$ according to:
\begin{equation}
\lambda=4\pi \left[8(1-C)(1+\nu)(\pi_{\phi\phi}/\gamma_{o/w}-1)\right]^{-1/2} r_s
\label{eq:theo:lambda}
\end{equation}

\section{Mohr-Coulomb criterion for 2D cohesive and frictional granular materials}\label{app:mohrcoulomb}

The Mohr Coulomb criterion for plasticity was derived in the past for 3D granular media \cite{Andreotti2013}. Here, it is recast in a two-dimensional geometry and writes 
\begin{equation}
(\sigma_1-\sigma_2)^2=\sin^2\delta \left(\sigma_1+\sigma_2+\frac{2c}{\tan \delta}\right)^2
\label{eq:MCapp}
\end{equation}
where $c$ is the cohesion and $\delta$ is the friction angle between two particles. At the wrinkling threshold, we find $\pi_{zz}>\pi_{\phi\phi}$ so that we identify $\sigma_{1,2}$ as follows: $\sigma_1=\pi_{zz}$ and $\sigma_2=\pi_{\phi\phi}$. Equation~\ref{eq:MCapp} becomes Equation~\ref{eq:pi:phi:phi}.\\

In the following, we offer a physical description of the cohesive term $c$ we measure. In granular materials, the cohesion $c$ is related to the friction coefficient between particles $\tan \delta$ and the internal stress arising from the attractive interactions between particles through :
\begin{equation}
c=\tan \delta \sigma_c
\label{eq:def:c}
\end{equation}
where $\sigma_c$ is the radial component of the Irwing Kirkwood tensor, or the contact stress. The purpose here is to link the contact stress to the attractive capillary forces between particles. The general equation for $\sigma_c$ writes:
\begin{equation}
\sigma_c=\frac{1}{S}\Sigma \left<f_x b_x\right>
\label{eq:def:sigmac}
\end{equation}
where $x$ is an arbitrary axis within the interface plane, $f_x$ the projection along $x$ of the force acting between two particles, and $b_x$ the projection along $x$ of the vector joining the centers of the two considered particles. Sum is made over the $N$ pairs of particles interacting on the surface $S$. From microscopy images in Fig.~\ref{fig:schema:montage}, the number of neighbors is approximately 6 per particle so that the number of contacts per particle is 3. Using the surface density of particles $C=\frac{\pi r_s^2}{S}$ as defined in Section~\ref{sec:exp}, the average number of contacts per unit surface writes $N/S=3\frac{C}{\pi r_s^2}$. The average distance between the center of two contacting particles is of order $2r_s$ so that $\left<f_xb_x\right>\sim r_sf_r$ where $f_r$ is the mean attractive force between particles.
Following models from the literature \cite{Danov:2010} accounting for capillary attractive forces arising from the pinning of contact lines at the particle interfaces, we offer to derive the interparticulate force according to:
\begin{equation}
f_r\simeq \epsilon \gamma_{o/w} \pi r_s
\label{eq:fr:epsilon}
\end{equation}
where $\epsilon$ is a numerical factor characterizing the non circularity of the oil/water contact line at the particle interface. Equations~\ref{eq:def:c}, \ref{eq:def:sigmac}, and \ref{eq:fr:epsilon} allow to relate the cohesion $c$ that we measure to the geometrical parameter $\epsilon$:
\begin{equation}
c=\tan \delta C\epsilon \gamma_{o/w}
\end{equation}
Taking $c=0.07 \gamma_{o/w}$ from Section~\ref{sec:mech:2D}, $C=0.86$ and $\delta=20^o$ \cite{Taran2006}, we find $\epsilon \simeq 0.4 $.



\begin{thebibliography}{0}%
\makeatletter
\providecommand \@ifxundefined [1]{%
 \@ifx{#1\undefined}
}%
\providecommand \@ifnum [1]{%
 \ifnum #1\expandafter \@firstoftwo
 \else \expandafter \@secondoftwo
 \fi
}%
\providecommand \@ifx [1]{%
 \ifx #1\expandafter \@firstoftwo
 \else \expandafter \@secondoftwo
 \fi
}%
\providecommand \natexlab [1]{#1}%
\providecommand \enquote  [1]{``#1''}%
\providecommand \bibnamefont  [1]{#1}%
\providecommand \bibfnamefont [1]{#1}%
\providecommand \citenamefont [1]{#1}%
\providecommand \href@noop [0]{\@secondoftwo}%
\providecommand \href [0]{\begingroup \@sanitize@url \@href}%
\providecommand \@href[1]{\@@startlink{#1}\@@href}%
\providecommand \@@href[1]{\endgroup#1\@@endlink}%
\providecommand \@sanitize@url [0]{\catcode `\\12\catcode `\$12\catcode
  `\&12\catcode `\#12\catcode `\^12\catcode `\_12\catcode `\%12\relax}%
\providecommand \@@startlink[1]{}%
\providecommand \@@endlink[0]{}%
\providecommand \url  [0]{\begingroup\@sanitize@url \@url }%
\providecommand \@url [1]{\endgroup\@href {#1}{\urlprefix }}%
\providecommand \urlprefix  [0]{URL }%
\providecommand \Eprint [0]{\href }%
\providecommand \doibase [0]{https://doi.org/}%
\providecommand \selectlanguage [0]{\@gobble}%
\providecommand \bibinfo  [0]{\@secondoftwo}%
\providecommand \bibfield  [0]{\@secondoftwo}%
\providecommand \translation [1]{[#1]}%
\providecommand \BibitemOpen [0]{}%
\providecommand \bibitemStop [0]{}%
\providecommand \bibitemNoStop [0]{.\EOS\space}%
\providecommand \EOS [0]{\spacefactor3000\relax}%
\providecommand \BibitemShut  [1]{\csname bibitem#1\endcsname}%
\let\auto@bib@innerbib\@empty
\end{thebibliography}%


\begin{thebibliography}{43}%
	\makeatletter
	\providecommand \@ifxundefined [1]{%
		\@ifx{#1\undefined}
	}%
	\providecommand \@ifnum [1]{%
		\ifnum #1\expandafter \@firstoftwo
		\else \expandafter \@secondoftwo
		\fi
	}%
	\providecommand \@ifx [1]{%
		\ifx #1\expandafter \@firstoftwo
		\else \expandafter \@secondoftwo
		\fi
	}%
	\providecommand \natexlab [1]{#1}%
	\providecommand \enquote  [1]{``#1''}%
	\providecommand \bibnamefont  [1]{#1}%
	\providecommand \bibfnamefont [1]{#1}%
	\providecommand \citenamefont [1]{#1}%
	\providecommand \href@noop [0]{\@secondoftwo}%
	\providecommand \href [0]{\begingroup \@sanitize@url \@href}%
	\providecommand \@href[1]{\@@startlink{#1}\@@href}%
	\providecommand \@@href[1]{\endgroup#1\@@endlink}%
	\providecommand \@sanitize@url [0]{\catcode `\\12\catcode `\$12\catcode
		`\&12\catcode `\#12\catcode `\^12\catcode `\_12\catcode `\%12\relax}%
	\providecommand \@@startlink[1]{}%
	\providecommand \@@endlink[0]{}%
	\providecommand \url  [0]{\begingroup\@sanitize@url \@url }%
	\providecommand \@url [1]{\endgroup\@href {#1}{\urlprefix }}%
	\providecommand \urlprefix  [0]{URL }%
	\providecommand \Eprint [0]{\href }%
	\providecommand \doibase [0]{https://doi.org/}%
	\providecommand \selectlanguage [0]{\@gobble}%
	\providecommand \bibinfo  [0]{\@secondoftwo}%
	\providecommand \bibfield  [0]{\@secondoftwo}%
	\providecommand \translation [1]{[#1]}%
	\providecommand \BibitemOpen [0]{}%
	\providecommand \bibitemStop [0]{}%
	\providecommand \bibitemNoStop [0]{.\EOS\space}%
	\providecommand \EOS [0]{\spacefactor3000\relax}%
	\providecommand \BibitemShut  [1]{\csname bibitem#1\endcsname}%
	\let\auto@bib@innerbib\@empty
	\bibitem [{\citenamefont {Pieranski}(1980)}]{Pieranski1980}%
	\BibitemOpen
	\bibfield  {author} {\bibinfo {author} {\bibfnamefont {P.}~\bibnamefont
			{Pieranski}},\ }\bibfield  {title} {\bibinfo {title} {Two-dimensional
			interfacial colloidal crystals},\ }\href
	{https://doi.org/10.1103/PhysRevLett.45.569} {\bibfield  {journal} {\bibinfo
			{journal} {Phys. Rev. Lett.}\ }\textbf {\bibinfo {volume} {45}},\ \bibinfo
		{pages} {569} (\bibinfo {year} {1980})}\BibitemShut {NoStop}%
	\bibitem [{\citenamefont {Chevalier}\ and\ \citenamefont
		{Bolzinger}(2013)}]{Chevalier2013}%
	\BibitemOpen
	\bibfield  {author} {\bibinfo {author} {\bibfnamefont {Y.}~\bibnamefont
			{Chevalier}}\ and\ \bibinfo {author} {\bibfnamefont {M.-A.}\ \bibnamefont
			{Bolzinger}},\ }\bibfield  {title} {\bibinfo {title} {Emulsions stabilized
			with solid nanoparticles: Pickering emulsions},\ }\href
	{https://doi.org/https://doi.org/10.1016/j.colsurfa.2013.02.054} {\bibfield
		{journal} {\bibinfo  {journal} {Colloids and Surfaces A: Physicochemical and
				Engineering Aspects}\ }\textbf {\bibinfo {volume} {439}},\ \bibinfo {pages}
		{23} (\bibinfo {year} {2013})}\BibitemShut {NoStop}%
	\bibitem [{\citenamefont {Gai}\ \emph {et~al.}(2017)\citenamefont {Gai},
		\citenamefont {Kim}, \citenamefont {Pan},\ and\ \citenamefont
		{Tang}}]{gai_amphiphilic_2017}%
	\BibitemOpen
	\bibfield  {author} {\bibinfo {author} {\bibfnamefont {Y.}~\bibnamefont
			{Gai}}, \bibinfo {author} {\bibfnamefont {M.}~\bibnamefont {Kim}}, \bibinfo
		{author} {\bibfnamefont {M.}~\bibnamefont {Pan}},\ and\ \bibinfo {author}
		{\bibfnamefont {S.~K.~Y.}\ \bibnamefont {Tang}},\ }\bibfield  {title}
	{\bibinfo {title} {Amphiphilic nanoparticles suppress droplet break-up in a
			concentrated emulsion flowing through a narrow constriction},\ }\bibfield
	{journal} {\bibinfo  {journal} {Biomicrofluidics}\ }\textbf {\bibinfo
		{volume} {11}},\ \href {https://doi.org/10.1063/1.4985158}
	{10.1063/1.4985158} (\bibinfo {year} {2017})\BibitemShut {NoStop}%
	\bibitem [{\citenamefont {Lagarde}\ and\ \citenamefont
		{Protiere}(2020)}]{Lagarde:protiere:2020}%
	\BibitemOpen
	\bibfield  {author} {\bibinfo {author} {\bibfnamefont {A.}~\bibnamefont
			{Lagarde}}\ and\ \bibinfo {author} {\bibfnamefont {S.}~\bibnamefont
			{Protiere}},\ }\bibfield  {title} {\bibinfo {title} {Probing the erosion and
			cohesion of a granular raft in motion},\ }\bibfield  {journal} {\bibinfo
		{journal} {Physical Review Fluids}\ }\textbf {\bibinfo {volume} {5}},\ \href
	{https://doi.org/10.1103/PhysRevFluids.5.044003}
	{10.1103/PhysRevFluids.5.044003} (\bibinfo {year} {2020})\BibitemShut
	{NoStop}%
	\bibitem [{\citenamefont {Smith}\ and\ \citenamefont {{van de
				Ven}}(1985)}]{Smith1985}%
	\BibitemOpen
	\bibfield  {author} {\bibinfo {author} {\bibfnamefont {P.}~\bibnamefont
			{Smith}}\ and\ \bibinfo {author} {\bibfnamefont {T.}~\bibnamefont {{van de
					Ven}}},\ }\bibfield  {title} {\bibinfo {title} {Shear-induced deformation and
			rupture of suspended solid/liquid clusters},\ }\href
	{https://doi.org/https://doi.org/10.1016/0166-6622(85)80071-8} {\bibfield
		{journal} {\bibinfo  {journal} {Colloids and Surfaces}\ }\textbf {\bibinfo
			{volume} {15}},\ \bibinfo {pages} {191} (\bibinfo {year} {1985})}\BibitemShut
	{NoStop}%
	\bibitem [{\citenamefont {Mehrabian}\ \emph {et~al.}(2015)\citenamefont
		{Mehrabian}, \citenamefont {Bussmann},\ and\ \citenamefont
		{Acosta}}]{Mehrabian2015}%
	\BibitemOpen
	\bibfield  {author} {\bibinfo {author} {\bibfnamefont {S.}~\bibnamefont
			{Mehrabian}}, \bibinfo {author} {\bibfnamefont {M.}~\bibnamefont
			{Bussmann}},\ and\ \bibinfo {author} {\bibfnamefont {E.}~\bibnamefont
			{Acosta}},\ }\bibfield  {title} {\bibinfo {title} {Breakup of high solid
			volume fraction oil–particle cluster in simple shear flow},\ }\href
	{https://doi.org/https://doi.org/10.1016/j.colsurfa.2015.06.054} {\bibfield
		{journal} {\bibinfo  {journal} {Colloids and Surfaces A: Physicochemical and
				Engineering Aspects}\ }\textbf {\bibinfo {volume} {483}},\ \bibinfo {pages}
		{25} (\bibinfo {year} {2015})}\BibitemShut {NoStop}%
	\bibitem [{\citenamefont {Perazzo}\ \emph {et~al.}(2018)\citenamefont
		{Perazzo}, \citenamefont {Tomaiuolo}, \citenamefont {Preziosi},\ and\
		\citenamefont {Guido}}]{Perazzo2018}%
	\BibitemOpen
	\bibfield  {author} {\bibinfo {author} {\bibfnamefont {A.}~\bibnamefont
			{Perazzo}}, \bibinfo {author} {\bibfnamefont {G.}~\bibnamefont {Tomaiuolo}},
		\bibinfo {author} {\bibfnamefont {V.}~\bibnamefont {Preziosi}},\ and\
		\bibinfo {author} {\bibfnamefont {S.}~\bibnamefont {Guido}},\ }\bibfield
	{title} {\bibinfo {title} {Emulsions in porous media: From single droplet
			behavior to applications for oil recovery},\ }\href
	{https://doi.org/https://doi.org/10.1016/j.cis.2018.03.002} {\bibfield
		{journal} {\bibinfo  {journal} {Advances in Colloid and Interface Science}\
		}\textbf {\bibinfo {volume} {256}},\ \bibinfo {pages} {305} (\bibinfo {year}
		{2018})}\BibitemShut {NoStop}%
	\bibitem [{\citenamefont {Kokal}(2005)}]{kokal2005}%
	\BibitemOpen
	\bibfield  {author} {\bibinfo {author} {\bibfnamefont {S.}~\bibnamefont
			{Kokal}},\ }\bibfield  {title} {\bibinfo {title} {Crude-oil emulsions: A
			state-of-the-art review},\ }\href {https://doi.org/10.2118/77497-PA}
	{\bibfield  {journal} {\bibinfo  {journal} {SPE Production \& Facilities}\
		}\textbf {\bibinfo {volume} {20}},\ \bibinfo {pages} {5} (\bibinfo {year}
		{2005})},\ \bibinfo {note} {2002 SPE Annual Technical Conference and
		Exhibition, SAN ANTONIO, TX, SEP 29-OCT 02, 2002}\BibitemShut {NoStop}%
	\bibitem [{\citenamefont {Dressaire}\ and\ \citenamefont
		{Sauret}(2017)}]{Sauret2017}%
	\BibitemOpen
	\bibfield  {author} {\bibinfo {author} {\bibfnamefont {E.}~\bibnamefont
			{Dressaire}}\ and\ \bibinfo {author} {\bibfnamefont {A.}~\bibnamefont
			{Sauret}},\ }\bibfield  {title} {\bibinfo {title} {Clogging of microfluidic
			systems},\ }\href {https://doi.org/10.1039/C6SM01879C} {\bibfield  {journal}
		{\bibinfo  {journal} {Soft Matter}\ }\textbf {\bibinfo {volume} {13}},\
		\bibinfo {pages} {37} (\bibinfo {year} {2017})}\BibitemShut {NoStop}%
	\bibitem [{\citenamefont {Wyss}\ \emph {et~al.}(2010)\citenamefont {Wyss},
		\citenamefont {Franke}, \citenamefont {Mele},\ and\ \citenamefont
		{Weitz}}]{wyss_capillary_2010}%
	\BibitemOpen
	\bibfield  {author} {\bibinfo {author} {\bibfnamefont {H.~M.}\ \bibnamefont
			{Wyss}}, \bibinfo {author} {\bibfnamefont {T.}~\bibnamefont {Franke}},
		\bibinfo {author} {\bibfnamefont {E.}~\bibnamefont {Mele}},\ and\ \bibinfo
		{author} {\bibfnamefont {D.~A.}\ \bibnamefont {Weitz}},\ }\bibfield  {title}
	{\bibinfo {title} {Capillary micromechanics: {Measuring} the elasticity of
			microscopic soft objects},\ }\href {https://doi.org/10.1039/c003344h}
	{\bibfield  {journal} {\bibinfo  {journal} {Soft Matter}\ }\textbf {\bibinfo
			{volume} {6}},\ \bibinfo {pages} {4550} (\bibinfo {year} {2010})}\BibitemShut
	{NoStop}%
	\bibitem [{\citenamefont {Fiddes}\ \emph {et~al.}(2009)\citenamefont {Fiddes},
		\citenamefont {Chan}, \citenamefont {Wyss}, \citenamefont {Simmons},
		\citenamefont {Kumacheva},\ and\ \citenamefont
		{Wheeler}}]{fiddes_augmenting_2009}%
	\BibitemOpen
	\bibfield  {author} {\bibinfo {author} {\bibfnamefont {L.~K.}\ \bibnamefont
			{Fiddes}}, \bibinfo {author} {\bibfnamefont {H.~K.~C.}\ \bibnamefont {Chan}},
		\bibinfo {author} {\bibfnamefont {K.}~\bibnamefont {Wyss}}, \bibinfo {author}
		{\bibfnamefont {C.~A.}\ \bibnamefont {Simmons}}, \bibinfo {author}
		{\bibfnamefont {E.}~\bibnamefont {Kumacheva}},\ and\ \bibinfo {author}
		{\bibfnamefont {A.~R.}\ \bibnamefont {Wheeler}},\ }\bibfield  {title}
	{\bibinfo {title} {Augmenting microgel flow via receptor-ligand binding in
			the constrained geometries of microchannels},\ }\href
	{https://doi.org/10.1039/b807106c} {\bibfield  {journal} {\bibinfo  {journal}
			{Lab on a Chip}\ }\textbf {\bibinfo {volume} {9}},\ \bibinfo {pages} {286}
		(\bibinfo {year} {2009})}\BibitemShut {NoStop}%
	\bibitem [{\citenamefont {Leclerc}\ \emph {et~al.}(2012)\citenamefont
		{Leclerc}, \citenamefont {Kinoshita}, \citenamefont {Fujii},\ and\
		\citenamefont {Barthès-Biesel}}]{leclerc_transient_2012}%
	\BibitemOpen
	\bibfield  {author} {\bibinfo {author} {\bibfnamefont {E.}~\bibnamefont
			{Leclerc}}, \bibinfo {author} {\bibfnamefont {H.}~\bibnamefont {Kinoshita}},
		\bibinfo {author} {\bibfnamefont {T.}~\bibnamefont {Fujii}},\ and\ \bibinfo
		{author} {\bibfnamefont {D.}~\bibnamefont {Barthès-Biesel}},\ }\bibfield
	{title} {\bibinfo {title} {Transient flow of microcapsules through
			convergent–divergent microchannels},\ }\href
	{https://doi.org/10.1007/s10404-011-0907-1} {\bibfield  {journal} {\bibinfo
			{journal} {Microfluidics and Nanofluidics}\ }\textbf {\bibinfo {volume}
			{12}},\ \bibinfo {pages} {761} (\bibinfo {year} {2012})}\BibitemShut
	{NoStop}%
	\bibitem [{\citenamefont {Dawson}\ \emph {et~al.}(2015)\citenamefont {Dawson},
		\citenamefont {Häner},\ and\ \citenamefont {Juel}}]{dawson_extreme_2015}%
	\BibitemOpen
	\bibfield  {author} {\bibinfo {author} {\bibfnamefont {G.}~\bibnamefont
			{Dawson}}, \bibinfo {author} {\bibfnamefont {E.}~\bibnamefont {Häner}},\
		and\ \bibinfo {author} {\bibfnamefont {A.}~\bibnamefont {Juel}},\ }\bibfield
	{title} {\bibinfo {title} {Extreme {Deformation} of {Capsules} and {Bubbles}
			{Flowing} through a {Localised} {Constriction}},\ }\href
	{https://doi.org/https://doi.org/10.1016/j.piutam.2015.03.004} {\bibfield
		{journal} {\bibinfo  {journal} {Procedia IUTAM}\ }\textbf {\bibinfo {volume}
			{16}},\ \bibinfo {pages} {22} (\bibinfo {year} {2015})}\BibitemShut {NoStop}%
	\bibitem [{\citenamefont {Rorai}\ \emph {et~al.}(2015)\citenamefont {Rorai},
		\citenamefont {Touchard}, \citenamefont {Zhu},\ and\ \citenamefont
		{Brandt}}]{rorai_motion_2015}%
	\BibitemOpen
	\bibfield  {author} {\bibinfo {author} {\bibfnamefont {C.}~\bibnamefont
			{Rorai}}, \bibinfo {author} {\bibfnamefont {A.}~\bibnamefont {Touchard}},
		\bibinfo {author} {\bibfnamefont {L.}~\bibnamefont {Zhu}},\ and\ \bibinfo
		{author} {\bibfnamefont {L.}~\bibnamefont {Brandt}},\ }\bibfield  {title}
	{\bibinfo {title} {Motion of an elastic capsule in a constricted
			microchannel},\ }\bibfield  {journal} {\bibinfo  {journal} {European Physical
			Journal E}\ }\textbf {\bibinfo {volume} {38}},\ \href
	{https://doi.org/10.1140/epje/i2015-15049-8} {10.1140/epje/i2015-15049-8}
	(\bibinfo {year} {2015})\BibitemShut {NoStop}%
	\bibitem [{\citenamefont {Luo}\ and\ \citenamefont
		{Bai}(2019)}]{luo_solute_2019}%
	\BibitemOpen
	\bibfield  {author} {\bibinfo {author} {\bibfnamefont {Z.~Y.}\ \bibnamefont
			{Luo}}\ and\ \bibinfo {author} {\bibfnamefont {B.~F.}\ \bibnamefont {Bai}},\
	}\bibfield  {title} {\bibinfo {title} {Solute release from an elastic capsule
			flowing through a microfluidic channel constriction},\ }\bibfield  {journal}
	{\bibinfo  {journal} {Physics of Fluids}\ }\textbf {\bibinfo {volume} {31}},\
	\href {https://doi.org/10.1063/1.5129413} {10.1063/1.5129413} (\bibinfo
	{year} {2019})\BibitemShut {NoStop}%
	\bibitem [{\citenamefont {Haner}\ \emph {et~al.}(2021)\citenamefont {Haner},
		\citenamefont {Vesperini}, \citenamefont {Salsac}, \citenamefont {Le~Goff},\
		and\ \citenamefont {Juel}}]{haner_sorting_2021}%
	\BibitemOpen
	\bibfield  {author} {\bibinfo {author} {\bibfnamefont {E.}~\bibnamefont
			{Haner}}, \bibinfo {author} {\bibfnamefont {D.}~\bibnamefont {Vesperini}},
		\bibinfo {author} {\bibfnamefont {A.-V.}\ \bibnamefont {Salsac}}, \bibinfo
		{author} {\bibfnamefont {A.}~\bibnamefont {Le~Goff}},\ and\ \bibinfo {author}
		{\bibfnamefont {A.}~\bibnamefont {Juel}},\ }\bibfield  {title} {\bibinfo
		{title} {Sorting of capsules according to their stiffness: from principle to
			application},\ }\href {https://doi.org/10.1039/d0sm02249g} {\bibfield
		{journal} {\bibinfo  {journal} {Soft Matter}\ }\textbf {\bibinfo {volume}
			{17}},\ \bibinfo {pages} {3722} (\bibinfo {year} {2021})}\BibitemShut
	{NoStop}%
	\bibitem [{\citenamefont {Leopercio}\ \emph {et~al.}(2021)\citenamefont
		{Leopercio}, \citenamefont {Michelon},\ and\ \citenamefont
		{Carvalho}}]{leopercio_deformation_2021}%
	\BibitemOpen
	\bibfield  {author} {\bibinfo {author} {\bibfnamefont {B.~C.}\ \bibnamefont
			{Leopercio}}, \bibinfo {author} {\bibfnamefont {M.}~\bibnamefont
			{Michelon}},\ and\ \bibinfo {author} {\bibfnamefont {M.~S.}\ \bibnamefont
			{Carvalho}},\ }\bibfield  {title} {\bibinfo {title} {Deformation and rupture
			of microcapsules flowing through constricted capillary},\ }\bibfield
	{journal} {\bibinfo  {journal} {Scientific Reports}\ }\textbf {\bibinfo
		{volume} {11}},\ \href {https://doi.org/10.1038/s41598-021-86833-8}
	{10.1038/s41598-021-86833-8} (\bibinfo {year} {2021})\BibitemShut {NoStop}%
	\bibitem [{\citenamefont {Fai}\ \emph {et~al.}(2017)\citenamefont {Fai},
		\citenamefont {Kusters}, \citenamefont {Harting}, \citenamefont {Rycroft},\
		and\ \citenamefont {Mahadevan}}]{fai_active_2017}%
	\BibitemOpen
	\bibfield  {author} {\bibinfo {author} {\bibfnamefont {T.~G.}\ \bibnamefont
			{Fai}}, \bibinfo {author} {\bibfnamefont {R.}~\bibnamefont {Kusters}},
		\bibinfo {author} {\bibfnamefont {J.}~\bibnamefont {Harting}}, \bibinfo
		{author} {\bibfnamefont {C.~H.}\ \bibnamefont {Rycroft}},\ and\ \bibinfo
		{author} {\bibfnamefont {L.}~\bibnamefont {Mahadevan}},\ }\bibfield  {title}
	{\bibinfo {title} {Active elastohydrodynamics of vesicles in narrow blind
			constrictions},\ }\bibfield  {journal} {\bibinfo  {journal} {Physical Review
			Fluids}\ }\textbf {\bibinfo {volume} {2}},\ \href
	{https://doi.org/10.1103/PhysRevFluids.2.113601}
	{10.1103/PhysRevFluids.2.113601} (\bibinfo {year} {2017})\BibitemShut
	{NoStop}%
	\bibitem [{\citenamefont {Park}\ and\ \citenamefont
		{Fai}(2020)}]{park_dynamics_2020}%
	\BibitemOpen
	\bibfield  {author} {\bibinfo {author} {\bibfnamefont {Y.}~\bibnamefont
			{Park}}\ and\ \bibinfo {author} {\bibfnamefont {T.~G.}\ \bibnamefont {Fai}},\
	}\bibfield  {title} {\bibinfo {title} {Dynamics of {Vesicles} {Driven} {Into}
			{Closed} {Constrictions} by {Molecular} {Motors}},\ }\bibfield  {journal}
	{\bibinfo  {journal} {Bulletin of Mathematical Biology}\ }\textbf {\bibinfo
		{volume} {82}},\ \href {https://doi.org/10.1007/s11538-020-00820-0}
	{10.1007/s11538-020-00820-0} (\bibinfo {year} {2020})\BibitemShut {NoStop}%
	\bibitem [{\citenamefont {Prevot}\ \emph {et~al.}(2003)\citenamefont {Prevot},
		\citenamefont {Cordeiro}, \citenamefont {Sukhorukov}, \citenamefont {Lvov},
		\citenamefont {Besser},\ and\ \citenamefont {Mohwald}}]{prevot_design_2003}%
	\BibitemOpen
	\bibfield  {author} {\bibinfo {author} {\bibfnamefont {M.}~\bibnamefont
			{Prevot}}, \bibinfo {author} {\bibfnamefont {A.}~\bibnamefont {Cordeiro}},
		\bibinfo {author} {\bibfnamefont {G.}~\bibnamefont {Sukhorukov}}, \bibinfo
		{author} {\bibfnamefont {Y.}~\bibnamefont {Lvov}}, \bibinfo {author}
		{\bibfnamefont {R.}~\bibnamefont {Besser}},\ and\ \bibinfo {author}
		{\bibfnamefont {H.}~\bibnamefont {Mohwald}},\ }\bibfield  {title} {\bibinfo
		{title} {Design of a microfluidic system to investigate the mechanical
			properties of layer-by-layer fabricated capsules},\ }\href
	{https://doi.org/10.1002/mame.200300205} {\bibfield  {journal} {\bibinfo
			{journal} {Macrmolecular Materials and Engineering}\ }\textbf {\bibinfo
			{volume} {288}},\ \bibinfo {pages} {915} (\bibinfo {year}
		{2003})}\BibitemShut {NoStop}%
	\bibitem [{\citenamefont {DoNascimento}\ \emph {et~al.}(2017)\citenamefont
		{DoNascimento}, \citenamefont {Avendano}, \citenamefont {Mehl}, \citenamefont
		{Moura}, \citenamefont {Carvalho},\ and\ \citenamefont
		{Duncanson}}]{do_nascimento_flow_2017}%
	\BibitemOpen
	\bibfield  {author} {\bibinfo {author} {\bibfnamefont {D.~F.}\ \bibnamefont
			{DoNascimento}}, \bibinfo {author} {\bibfnamefont {J.~A.}\ \bibnamefont
			{Avendano}}, \bibinfo {author} {\bibfnamefont {A.}~\bibnamefont {Mehl}},
		\bibinfo {author} {\bibfnamefont {M.~J.~B.}\ \bibnamefont {Moura}}, \bibinfo
		{author} {\bibfnamefont {M.~S.}\ \bibnamefont {Carvalho}},\ and\ \bibinfo
		{author} {\bibfnamefont {W.~J.}\ \bibnamefont {Duncanson}},\ }\bibfield
	{title} {\bibinfo {title} {Flow of {Tunable} {Elastic} {Microcapsules}
			through {Constrictions}},\ }\bibfield  {journal} {\bibinfo  {journal}
		{Scientific Reports}\ }\textbf {\bibinfo {volume} {7}},\ \href
	{https://doi.org/10.1038/s41598-017-11950-2} {10.1038/s41598-017-11950-2}
	(\bibinfo {year} {2017})\BibitemShut {NoStop}%
	\bibitem [{\citenamefont {O'Connell}\ \emph {et~al.}(2019)\citenamefont
		{O'Connell}, \citenamefont {Lu}, \citenamefont {Browne},\ and\ \citenamefont
		{Datta}}]{oconnell_cooperative_2019}%
	\BibitemOpen
	\bibfield  {author} {\bibinfo {author} {\bibfnamefont {M.~G.}\ \bibnamefont
			{O'Connell}}, \bibinfo {author} {\bibfnamefont {N.~B.}\ \bibnamefont {Lu}},
		\bibinfo {author} {\bibfnamefont {C.~A.}\ \bibnamefont {Browne}},\ and\
		\bibinfo {author} {\bibfnamefont {S.~S.}\ \bibnamefont {Datta}},\ }\bibfield
	{title} {\bibinfo {title} {Cooperative size sorting of deformable particles
			in porous media},\ }\href {https://doi.org/10.1039/c9sm00300b} {\bibfield
		{journal} {\bibinfo  {journal} {Soft Matter}\ }\textbf {\bibinfo {volume}
			{15}},\ \bibinfo {pages} {3620} (\bibinfo {year} {2019})}\BibitemShut
	{NoStop}%
	\bibitem [{\citenamefont {Fiddes}\ \emph {et~al.}(2007)\citenamefont {Fiddes},
		\citenamefont {Young}, \citenamefont {Kumacheva},\ and\ \citenamefont
		{Wheeler}}]{fiddes_flow_2007}%
	\BibitemOpen
	\bibfield  {author} {\bibinfo {author} {\bibfnamefont {L.~K.}\ \bibnamefont
			{Fiddes}}, \bibinfo {author} {\bibfnamefont {E.~W.~K.}\ \bibnamefont
			{Young}}, \bibinfo {author} {\bibfnamefont {E.}~\bibnamefont {Kumacheva}},\
		and\ \bibinfo {author} {\bibfnamefont {A.~R.}\ \bibnamefont {Wheeler}},\
	}\bibfield  {title} {\bibinfo {title} {Flow of microgel capsules through
			topographically patterned microchannels},\ }\href
	{https://doi.org/10.1039/b703297h} {\bibfield  {journal} {\bibinfo  {journal}
			{Lab on a Chip}\ }\textbf {\bibinfo {volume} {7}},\ \bibinfo {pages} {863}
		(\bibinfo {year} {2007})}\BibitemShut {NoStop}%
	\bibitem [{\citenamefont {Mulligan}\ and\ \citenamefont
		{Rothstein}(2011)}]{Mulligan2011}%
	\BibitemOpen
	\bibfield  {author} {\bibinfo {author} {\bibfnamefont {M.~K.}\ \bibnamefont
			{Mulligan}}\ and\ \bibinfo {author} {\bibfnamefont {J.~P.}\ \bibnamefont
			{Rothstein}},\ }\bibfield  {title} {\bibinfo {title} {Deformation and breakup
			of micro- and nanoparticle stabilized droplets in microfluidic extensional
			flows},\ }\href {https://doi.org/10.1021/la201523r} {\bibfield  {journal}
		{\bibinfo  {journal} {Langmuir}\ }\textbf {\bibinfo {volume} {27}},\ \bibinfo
		{pages} {9760} (\bibinfo {year} {2011})},\ \bibinfo {note} {pMID: 21732665},\
	\Eprint {https://arxiv.org/abs/https://doi.org/10.1021/la201523r}
	{https://doi.org/10.1021/la201523r} \BibitemShut {NoStop}%
	\bibitem [{\citenamefont {Liu}\ \emph {et~al.}(2021)\citenamefont {Liu},
		\citenamefont {Sun},\ and\ \citenamefont {Santamarina}}]{Liu2021}%
	\BibitemOpen
	\bibfield  {author} {\bibinfo {author} {\bibfnamefont {Q.}~\bibnamefont
			{Liu}}, \bibinfo {author} {\bibfnamefont {Z.}~\bibnamefont {Sun}},\ and\
		\bibinfo {author} {\bibfnamefont {J.~C.}\ \bibnamefont {Santamarina}},\
	}\bibfield  {title} {\bibinfo {title} {Self-assembled nanoparticle-coated
			interfaces: Capillary pressure, shell formation and buckling},\ }\href
	{https://doi.org/https://doi.org/10.1016/j.jcis.2020.07.110} {\bibfield
		{journal} {\bibinfo  {journal} {Journal of Colloid and Interface Science}\
		}\textbf {\bibinfo {volume} {581}},\ \bibinfo {pages} {251} (\bibinfo {year}
		{2021})}\BibitemShut {NoStop}%
	\bibitem [{\citenamefont {De~Soete}\ \emph {et~al.}(2021)\citenamefont
		{De~Soete}, \citenamefont {Delance}, \citenamefont {Passade-Boupat},
		\citenamefont {Levant}, \citenamefont {Verneuil}, \citenamefont {Lequeux},\
		and\ \citenamefont {Talini}}]{DeSoete2021}%
	\BibitemOpen
	\bibfield  {author} {\bibinfo {author} {\bibfnamefont {F.}~\bibnamefont
			{De~Soete}}, \bibinfo {author} {\bibfnamefont {L.}~\bibnamefont {Delance}},
		\bibinfo {author} {\bibfnamefont {N.}~\bibnamefont {Passade-Boupat}},
		\bibinfo {author} {\bibfnamefont {M.}~\bibnamefont {Levant}}, \bibinfo
		{author} {\bibfnamefont {E.}~\bibnamefont {Verneuil}}, \bibinfo {author}
		{\bibfnamefont {F.}~\bibnamefont {Lequeux}},\ and\ \bibinfo {author}
		{\bibfnamefont {L.}~\bibnamefont {Talini}},\ }\bibfield  {title} {\bibinfo
		{title} {{Passage of surfactant-laden and particle-laden drops through a
				contraction}},\ }\bibfield  {journal} {\bibinfo  {journal} {{Physical Review
				Fluids}}\ }\textbf {\bibinfo {volume} {{6}}},\ \href
	{https://doi.org/{10.1103/PhysRevFluids.6.093601}}
	{{10.1103/PhysRevFluids.6.093601}} (\bibinfo {year} {{2021}})\BibitemShut
	{NoStop}%
	\bibitem [{\citenamefont {Bretherton}(1961)}]{Bretherton1961}%
	\BibitemOpen
	\bibfield  {author} {\bibinfo {author} {\bibfnamefont {F.~P.}\ \bibnamefont
			{Bretherton}},\ }\bibfield  {title} {\bibinfo {title} {The motion of long
			bubbles in tubes},\ }\href {https://doi.org/10.1017/S0022112061000160}
	{\bibfield  {journal} {\bibinfo  {journal} {Journal of Fluild Mechanics}\
		}\textbf {\bibinfo {volume} {10}},\ \bibinfo {pages} {166} (\bibinfo {year}
		{1961})}\BibitemShut {NoStop}%
	\bibitem [{SMv({\natexlab{a}})}]{SMvideo:freeflow}%
	\BibitemOpen
	\href@noop {} {\bibinfo {title} {{See Supplemental Material at [URL will be
				inserted by publisher] for a movie showing a drop laden with $r_s=0.75~\mu$m
				silica particles at $Ca=1.10^{-2}$ : At the drop front, the naked drop is
				supported by a lubrication water film of Bretherton's thickness $h=1.5~\mu$m.
				A cohesive particle raft initially accumulates at the back and is later
				recalled at the drop front. No significant particles expulsion.}}}
	({\natexlab{a}})\BibitemShut {NoStop}%
	\bibitem [{SMv({\natexlab{b}})}]{SMvideo:wrinkles}%
	\BibitemOpen
	\href@noop {} {\bibinfo {title} {{See Supplemental Material at [URL will be
				inserted by publisher] for a movie showing a drop laden with $r_s=0.75~\mu$m
				silica particles at $Ca=3.10^{-4}$. The Bretherton's film thickness would be
				$h=0.2~\mu$m. However, oil wets the wall at the front. At the drop back, the
				particles raft is compressed and wrinkles develop in the longitudinal
				direction as the back enters the converging part of the pore.}}}
	({\natexlab{b}})\BibitemShut {NoStop}%
	\bibitem [{SMv({\natexlab{c}})}]{SMvideo:expulsion}%
	\BibitemOpen
	\href@noop {} {\bibinfo {title} {{See Supplemental Material at [URL will be
				inserted by publisher] for a movie showing a drop laden with $r_s=5~\mu$m
				silica particles at $Ca=3.10^{-3}$ : Particles accumulate at the back. No
				wrinkles are observed. All particles are expelled from the drop interface
				after the drop exits the pore.}}} ({\natexlab{c}})\BibitemShut {NoStop}%
	\bibitem [{\citenamefont {Balestra}\ \emph {et~al.}(2018)\citenamefont
		{Balestra}, \citenamefont {Zhu},\ and\ \citenamefont
		{Gallaire}}]{Balestra:2018}%
	\BibitemOpen
	\bibfield  {author} {\bibinfo {author} {\bibfnamefont {G.}~\bibnamefont
			{Balestra}}, \bibinfo {author} {\bibfnamefont {L.}~\bibnamefont {Zhu}},\ and\
		\bibinfo {author} {\bibfnamefont {F.}~\bibnamefont {Gallaire}},\ }\bibfield
	{title} {\bibinfo {title} {Viscous taylor droplets in axisymmetric and planar
			tubes: from bretherton's theory to empirical models},\ }\bibfield  {journal}
	{\bibinfo  {journal} {Microfluidics and Nanofluidics}\ }\textbf {\bibinfo
		{volume} {22}},\ \href {https://doi.org/10.1007/s10404-018-2084-y}
	{10.1007/s10404-018-2084-y} (\bibinfo {year} {2018})\BibitemShut {NoStop}%
	\bibitem [{\citenamefont {Aussillous}\ and\ \citenamefont
		{Quere}(2000)}]{Aussilous2000}%
	\BibitemOpen
	\bibfield  {author} {\bibinfo {author} {\bibfnamefont {P.}~\bibnamefont
			{Aussillous}}\ and\ \bibinfo {author} {\bibfnamefont {D.}~\bibnamefont
			{Quere}},\ }\bibfield  {title} {\bibinfo {title} {Quick deposition of a fluid
			on the wall of a tube},\ }\href {https://doi.org/10.1063/1.1289396}
	{\bibfield  {journal} {\bibinfo  {journal} {Physics of Fluids}\ }\textbf
		{\bibinfo {volume} {12}},\ \bibinfo {pages} {2367} (\bibinfo {year}
		{2000})}\BibitemShut {NoStop}%
	\bibitem [{\citenamefont {Yu}\ \emph {et~al.}(2017)\citenamefont {Yu},
		\citenamefont {Khodaparast},\ and\ \citenamefont {Stone}}]{stone_2017}%
	\BibitemOpen
	\bibfield  {author} {\bibinfo {author} {\bibfnamefont {Y.~E.}\ \bibnamefont
			{Yu}}, \bibinfo {author} {\bibfnamefont {S.}~\bibnamefont {Khodaparast}},\
		and\ \bibinfo {author} {\bibfnamefont {H.~A.}\ \bibnamefont {Stone}},\
	}\bibfield  {title} {\bibinfo {title} {Armoring confined bubbles in the flow
			of colloidal suspensions},\ }\href {https://doi.org/10.1039/c6sm02585d}
	{\bibfield  {journal} {\bibinfo  {journal} {Soft Matter}\ }\textbf {\bibinfo
			{volume} {13}},\ \bibinfo {pages} {2857} (\bibinfo {year}
		{2017})}\BibitemShut {NoStop}%
	\bibitem [{\citenamefont {Rondepierre}\ \emph {et~al.}(2021)\citenamefont
		{Rondepierre}, \citenamefont {De~Soete}, \citenamefont {Passade-Boupat},
		\citenamefont {Lequeux}, \citenamefont {Talini}, \citenamefont {Limat},\ and\
		\citenamefont {Verneuil}}]{Rondepierre2021}%
	\BibitemOpen
	\bibfield  {author} {\bibinfo {author} {\bibfnamefont {G.}~\bibnamefont
			{Rondepierre}}, \bibinfo {author} {\bibfnamefont {F.}~\bibnamefont
			{De~Soete}}, \bibinfo {author} {\bibfnamefont {N.}~\bibnamefont
			{Passade-Boupat}}, \bibinfo {author} {\bibfnamefont {F.}~\bibnamefont
			{Lequeux}}, \bibinfo {author} {\bibfnamefont {L.}~\bibnamefont {Talini}},
		\bibinfo {author} {\bibfnamefont {L.}~\bibnamefont {Limat}},\ and\ \bibinfo
		{author} {\bibfnamefont {E.}~\bibnamefont {Verneuil}},\ }\bibfield  {title}
	{\bibinfo {title} {Dramatic slowing down of oil/water/silica contact line
			dynamics driven by cationic surfactant adsorption on the solid},\ }\href
	{https://doi.org/10.1021/acs.langmuir.0c02746} {\bibfield  {journal}
		{\bibinfo  {journal} {Langmuir}\ }\textbf {\bibinfo {volume} {37}},\ \bibinfo
		{pages} {1662} (\bibinfo {year} {2021})},\ \bibinfo {note} {pMID: 33502209},\
	\Eprint {https://arxiv.org/abs/https://doi.org/10.1021/acs.langmuir.0c02746}
	{https://doi.org/10.1021/acs.langmuir.0c02746} \BibitemShut {NoStop}%
	\bibitem [{\citenamefont {Jambon-Puillet}\ \emph {et~al.}(2017)\citenamefont
		{Jambon-Puillet}, \citenamefont {Josserand},\ and\ \citenamefont
		{Proti\`ere}}]{Jambon2017}%
	\BibitemOpen
	\bibfield  {author} {\bibinfo {author} {\bibfnamefont {E.}~\bibnamefont
			{Jambon-Puillet}}, \bibinfo {author} {\bibfnamefont {C.}~\bibnamefont
			{Josserand}},\ and\ \bibinfo {author} {\bibfnamefont {S.}~\bibnamefont
			{Proti\`ere}},\ }\bibfield  {title} {\bibinfo {title} {Wrinkles, folds, and
			plasticity in granular rafts},\ }\href
	{https://doi.org/10.1103/PhysRevMaterials.1.042601} {\bibfield  {journal}
		{\bibinfo  {journal} {Phys. Rev. Materials}\ }\textbf {\bibinfo {volume}
			{1}},\ \bibinfo {pages} {042601} (\bibinfo {year} {2017})}\BibitemShut
	{NoStop}%
	\bibitem [{\citenamefont {Garbin}(2019)}]{Garbin2019}%
	\BibitemOpen
	\bibfield  {author} {\bibinfo {author} {\bibfnamefont {V.}~\bibnamefont
			{Garbin}},\ }\bibfield  {title} {\bibinfo {title} {Collapse mechanisms and
			extreme deformation of particle-laden interfaces},\ }\href
	{https://doi.org/https://doi.org/10.1016/j.cocis.2019.02.007} {\bibfield
		{journal} {\bibinfo  {journal} {Current Opinion in Colloid and Interface
				Science}\ }\textbf {\bibinfo {volume} {39}},\ \bibinfo {pages} {202}
		(\bibinfo {year} {2019})},\ \bibinfo {note} {special Topic Section:
		Outstanding Young Researchers in Colloid and Interface Science}\BibitemShut
	{NoStop}%
	\bibitem [{\citenamefont {Pitois}\ and\ \citenamefont
		{Rouyer}(2019)}]{Pitois2019}%
	\BibitemOpen
	\bibfield  {author} {\bibinfo {author} {\bibfnamefont {O.}~\bibnamefont
			{Pitois}}\ and\ \bibinfo {author} {\bibfnamefont {F.}~\bibnamefont
			{Rouyer}},\ }\bibfield  {title} {\bibinfo {title} {Rheology of particulate
			rafts, films, and foams},\ }\href
	{https://doi.org/https://doi.org/10.1016/j.cocis.2019.05.004} {\bibfield
		{journal} {\bibinfo  {journal} {Current Opinion in Colloid and Interface
				Science}\ }\textbf {\bibinfo {volume} {43}},\ \bibinfo {pages} {125}
		(\bibinfo {year} {2019})},\ \bibinfo {note} {rheology}\BibitemShut {NoStop}%
	\bibitem [{\citenamefont {Vella}\ \emph {et~al.}(2004)\citenamefont {Vella},
		\citenamefont {Aussillous},\ and\ \citenamefont {Mahadevan}}]{Vella2004}%
	\BibitemOpen
	\bibfield  {author} {\bibinfo {author} {\bibfnamefont {D.}~\bibnamefont
			{Vella}}, \bibinfo {author} {\bibfnamefont {P.}~\bibnamefont {Aussillous}},\
		and\ \bibinfo {author} {\bibfnamefont {L.}~\bibnamefont {Mahadevan}},\
	}\bibfield  {title} {\bibinfo {title} {Elasticity of an interfacial particle
			raft},\ }\href {https://doi.org/10.1209/epl/i2004-10202-x} {\bibfield
		{journal} {\bibinfo  {journal} {Europhysics Letters ({EPL})}\ }\textbf
		{\bibinfo {volume} {68}},\ \bibinfo {pages} {212} (\bibinfo {year}
		{2004})}\BibitemShut {NoStop}%
	\bibitem [{\citenamefont {Protiere}\ \emph {et~al.}(2017)\citenamefont
		{Protiere}, \citenamefont {Josserand}, \citenamefont {Aristoff},
		\citenamefont {Stone},\ and\ \citenamefont {Abkarian}}]{Protiere:2017}%
	\BibitemOpen
	\bibfield  {author} {\bibinfo {author} {\bibfnamefont {S.}~\bibnamefont
			{Protiere}}, \bibinfo {author} {\bibfnamefont {C.}~\bibnamefont {Josserand}},
		\bibinfo {author} {\bibfnamefont {J.~M.}\ \bibnamefont {Aristoff}}, \bibinfo
		{author} {\bibfnamefont {H.~A.}\ \bibnamefont {Stone}},\ and\ \bibinfo
		{author} {\bibfnamefont {M.}~\bibnamefont {Abkarian}},\ }\bibfield  {title}
	{\bibinfo {title} {Sinking a granular raft},\ }\bibfield  {journal} {\bibinfo
		{journal} {Physical Review Letters}\ }\textbf {\bibinfo {volume} {118}},\
	\href {https://doi.org/10.1103/PhysRevLett.118.108001}
	{10.1103/PhysRevLett.118.108001} (\bibinfo {year} {2017})\BibitemShut
	{NoStop}%
	\bibitem [{\citenamefont {Andreotti}\ \emph {et~al.}(2013)\citenamefont
		{Andreotti}, \citenamefont {Forterre},\ and\ \citenamefont
		{Pouliquen}}]{Andreotti2013}%
	\BibitemOpen
	\bibfield  {author} {\bibinfo {author} {\bibfnamefont {B.}~\bibnamefont
			{Andreotti}}, \bibinfo {author} {\bibfnamefont {Y.}~\bibnamefont
			{Forterre}},\ and\ \bibinfo {author} {\bibfnamefont {O.}~\bibnamefont
			{Pouliquen}},\ }\bibinfo {title} {The granular solid: Plasticity},\ in\
	\href@noop {} {\emph {\bibinfo {booktitle} {Granular Media: Between Fluid and
				Solid}}}\ (\bibinfo  {publisher} {Cambridge University Press},\ \bibinfo
	{year} {2013})\ Chap.~\bibinfo {chapter} {4}, pp.\ \bibinfo {pages}
	{12--168}\BibitemShut {NoStop}%
	\bibitem [{\citenamefont {Danov}\ and\ \citenamefont
		{Kralchevsky}(2010)}]{Danov:2010}%
	\BibitemOpen
	\bibfield  {author} {\bibinfo {author} {\bibfnamefont {K.~D.}\ \bibnamefont
			{Danov}}\ and\ \bibinfo {author} {\bibfnamefont {P.~A.}\ \bibnamefont
			{Kralchevsky}},\ }\bibfield  {title} {\bibinfo {title} {Interaction between
			like-charged particles at a liquid interface: Electrostatic repulsion vs.
			electrocapillary attraction},\ }\href
	{https://doi.org/10.1016/j.jcis.2010.02.017} {\bibfield  {journal} {\bibinfo
			{journal} {Journal of Colloid and Interface Science}\ }\textbf {\bibinfo
			{volume} {345}},\ \bibinfo {pages} {505} (\bibinfo {year}
		{2010})}\BibitemShut {NoStop}%
	\bibitem [{\citenamefont {Benet}\ \emph {et~al.}(2018)\citenamefont {Benet},
		\citenamefont {Lostec}, \citenamefont {Pellegrino},\ and\ \citenamefont
		{Vernerey}}]{Benet:2018}%
	\BibitemOpen
	\bibfield  {author} {\bibinfo {author} {\bibfnamefont {E.}~\bibnamefont
			{Benet}}, \bibinfo {author} {\bibfnamefont {G.}~\bibnamefont {Lostec}},
		\bibinfo {author} {\bibfnamefont {J.}~\bibnamefont {Pellegrino}},\ and\
		\bibinfo {author} {\bibfnamefont {F.}~\bibnamefont {Vernerey}},\ }\bibfield
	{title} {\bibinfo {title} {Mechanical instability and percolation of
			deformable particles through porous networks},\ }\bibfield  {journal}
	{\bibinfo  {journal} {Physical Review E}\ }\textbf {\bibinfo {volume} {97}},\
	\href {https://doi.org/10.1103/PhysRevE.97.042607}
	{10.1103/PhysRevE.97.042607} (\bibinfo {year} {2018})\BibitemShut {NoStop}%
	\bibitem [{\citenamefont {Taran}\ \emph {et~al.}(2006)\citenamefont {Taran},
		\citenamefont {Donose}, \citenamefont {Vakarelski},\ and\ \citenamefont
		{Higashitani}}]{Taran2006}%
	\BibitemOpen
	\bibfield  {author} {\bibinfo {author} {\bibfnamefont {E.}~\bibnamefont
			{Taran}}, \bibinfo {author} {\bibfnamefont {B.~C.}\ \bibnamefont {Donose}},
		\bibinfo {author} {\bibfnamefont {I.~U.}\ \bibnamefont {Vakarelski}},\ and\
		\bibinfo {author} {\bibfnamefont {K.}~\bibnamefont {Higashitani}},\
	}\bibfield  {title} {\bibinfo {title} {ph dependence of friction forces
			between silica surfaces in solutions},\ }\href
	{https://doi.org/https://doi.org/10.1016/j.jcis.2005.10.038} {\bibfield
		{journal} {\bibinfo  {journal} {Journal of Colloid and Interface Science}\
		}\textbf {\bibinfo {volume} {297}},\ \bibinfo {pages} {199} (\bibinfo {year}
		{2006})}\BibitemShut {NoStop}%
\end{thebibliography}

%

\end{document}